\documentclass[a4paper,fleqn,usenatbib]{mnras}

\usepackage{newtxtext,newtxmath}

\usepackage[T1]{fontenc}
\usepackage{ae,aecompl}
\usepackage{threeparttable}
\usepackage{bm}

\usepackage{graphicx}	
\usepackage{amsmath}	
\usepackage{amssymb}	

\title[Shapes and alignments of clusters]
{Shapes and alignments of dark matter haloes and their brightest cluster galaxies in 39 strong lensing clusters}

\author[Taizo Okabe et al.]{Taizo Okabe$^{1}$,\thanks{E-mail: taizo.okabe@utap.phys.s.u-tokyo.ac.jp}
Masamune Oguri,$^{1,2,3}$
S{\'e}bastien Peirani,$^{4,5}$
Yasushi Suto,$^{1,2}$
\newauthor
Yohan Dubois,$^{5}$
Christophe Pichon,$^{5,6}$
Tetsu Kitayama,$^{7}$
Shin Sasaki,$^{8}$
\newauthor
and Takahiro Nishimichi$^{9,3}$
\\
$^{1}$Department of Physics, The University of Tokyo, 7-3-1 Hongo, Bunkyo-ku, Tokyo 113-0033, Japan\\
$^{2}$Research Center for the Early Universe, School of Science, The University of Tokyo, 7-3-1 Hongo, Bunkyo-ku, Tokyo, 113-0033, Japan\\
$^{3}$Kavli Institute for the Physics and Mathematics of the Universe (WPI), The University of Tokyo Institutes for Advanced Study, \\
The University of Tokyo, 5-1-5 Kashiwanoha, Kashiwa, Chiba 277-8583, Japan\\
$^{4}$Universit\'e C\^ote d'Azur, Observatoire de la C\^ote d'Azur, CNRS, Laboratoire Lagrange, Bd de l’Observatoire, CS 34229, 06304 Nice Cedex 4, France \\
$^{5}$Institut d'Astrophysique de Paris (UMR 7095: CNRS \& UPMC), 98 bis Bd Arago, 75014 Paris, France \\
$^{6}$Korea Institute for Advanced Study, 85 Hoegiro, Dongdaemun-gu, Seoul, 02455, Republic of Korea\\
$^{7}$Department of Physics, Toho University, Funabashi, 2-2-1 Miyama, Funabashi, Chiba 274-8510, Japan\\
$^{8}$Department of Physics, Tokyo Metropolitan University, 1-1 Minami-Osawa, Hachioji, Tokyo 192-0397, Japan\\
$^{9}$Center for Gravitational Physics, Yukawa Institute for Theoretical Physics, Kyoto University, Kyoto 606-8502, Japan\\
}

\date{Accepted XXX. Received YYY; in original form ZZZ}

\pubyear{2020}

\begin{document}
\label{firstpage}
\pagerange{\pageref{firstpage}--\pageref{lastpage}}
\maketitle

\begin{abstract}
We study shapes and alignments of 45 dark matter (DM) haloes and their brightest cluster galaxies (BCGs) using a sample of 39 massive clusters from Hubble Frontier Field (HFF), Cluster Lensing And Supernova survey with Hubble (CLASH), and Reionization Lensing Cluster Survey (RELICS). We measure shapes of the DM haloes by strong gravitational lensing, whereas BCG shapes are derived from their light profiles in {\it Hubble Space Telescope} images. 
Our measurements from a large sample of massive clusters presented here provide new constraints on dark matter and cluster astrophysics.
We find that DM haloes are on average highly elongated with the mean ellipticity of $0.482\pm 0.028$, and position angles of major axes of DM haloes and their BCGs tend to be aligned well with the mean value of alignment angles of $22.2\pm 3.9$~deg. We find that DM haloes in our sample are on average {\it more elongated} than their BCGs with the mean difference of their ellipticities of $0.11\pm 0.03$. In contrast, the Horizon-AGN cosmological hydrodynamical simulation predicts on average similar ellipticities between DM haloes and their central galaxies. While such a difference between the observations and the simulation may well be explained by the difference of their halo mass scales, other possibilities include the bias inherent to strong lensing measurements, limited knowledge of baryon physics, or a limitation of cold dark matter.  
\end{abstract}




\section{Introduction}

The standard cosmological model, dominated by cosmological constant and cold dark matter ($\Lambda$CDM), explains various observations over Mpc to Gpc scales, such as cosmic microwave background anisotropy \citep[e.g.,][]{2003ApJS..148..175S, 2016A&A...594A..13P}, the magnitude-redshift relation of Type Ia supernovae \citep[e.g.,][]{1999ApJ...517..565P, 1998AJ....116.1009R}, and baryon acoustic oscillations \citep[e.g.,][]{2005ApJ...633..560E}. Cosmological parameters of the $\Lambda$CDM model are determined very precisely from these observations. While the $\Lambda$CDM model has passed many observational tests, there still remain several challenges at small scales such as core-cusp and missing satellite problems \citep[e.g.,][for a review]{2017ARA&A..55..343B}. These challenges may point to an interesting possibility that the underlying assumption of the $\Lambda$CDM model has to be modified, including the modification of the nature of dark matter, although they might simply reflect a lack of our understanding of detailed baryon physics at small scales. Given their potential significance, it is important to confront $\Lambda$CDM model predictions at small scales ($\la 1$~Mpc) with a variety of observations.

Among others, galaxy clusters provide a useful means of testing the $\Lambda$CDM model, because their internal structure is mainly determined by the dynamics of dark matter and are less affected by detailed baryon physics at least as compared with galaxies for which effects of gas cooling and star formation on the internal structure are pronounced. An important characteristics of galaxy clusters is that they are highly non-spherical. There has been a number of observational studies that measure non-sphericities of galaxy clusters from member galaxy distribution \citep[e.g.,][]{1982A&A...107..338B, 2017NatAs...1E.157W}, X-ray surface brightness \citep[e.g.,][]{2008MNRAS.390.1562H, 2010ApJ...719.1926K}, gravitational lensing signal \citep[e.g.,][]{2010MNRAS.405.2215O, 2012MNRAS.420.3213O,2018ApJ...860..104U,2019arXiv191106333H}, and the Sunyaev Zel'dovich effect \citep[e.g.,][]{donahue}. Furthermore, $N$-body simulations based on the $\Lambda$CDM model also predict non-sphericities of galaxy cluster sized dark matter (DM) haloes \citep[e.g.,][]{2002ApJ...574..538J}. Since the degree of non-sphericities is sensitive to nature of dark matter \cite[e.g.,][]{2000ApJ...535L.103Y,2013MNRAS.430..105P}, comparison of observed non-sphericities of galaxy clusters serves as a useful complementary test of the $\Lambda$CDM model.

Previous comparisons against model predictions indicated that the observed non-sphericities of galaxy clusters are broadly consistent with the $\Lambda$CDM model. However, for more accurate test of the $\Lambda$CDM model both observations and theoretical model predictions have to be improved. While gravitational lensing provides a powerful tool to measure shapes of dark matter distributions directly, there is room for improvement in several ways. First, measurements of non-sphericities for individual clusters using weak lensing \citep[e.g.,][]{2010MNRAS.405.2215O, 2018ApJ...860..104U} are still noisy. Second, strong lensing allows us to measure shapes more accurately if there are sufficiently large number of multiple images \citep[e.g.,][]{2010MNRAS.404..325R,2012MNRAS.420.3213O}, but the sample size is not very large in practice ($N\sim 20$ including poorly constrained systems). Third, while stacked weak lensing enables us to measure the average shape of clusters very accurately \citep[e.g.,][]{2009ApJ...695.1446E,2016MNRAS.457.4135C,2017MNRAS.467.4131V, 2018MNRAS.475.2421S}, the interpretation of the observed signals is not easy because they depend on both non-sphericities of individual clusters and mis-alignments between prior directions of individual clusters used for stacking and their true orientations. In theoretical model predictions, even though the internal structure of clusters are relatively less affected by baryon physics, various baryon physics such as gas cooling, star formation, and feedback has non-negligible effects on shapes of dark matter distributions \citep[e.g.,][]{2004ApJ...611L..73K,suto17}. Thus the hydrodynamical simulation is required for accurate model predictions of cluster shapes. 

In this paper, we provide new measurements of shapes and orientations of 39 clusters by strong gravitational lensing. Our cluster sample is taken from recent three survey of clusters conduced with {\it Hubble Space Telescope (HST)}, Hubble Frontier Field \citep{2017ApJ...837...97L}, Cluster Lensing And Supernova survey with Hubble \citep{2012ApJS..199...25P}, and Reionization Lensing Cluster Survey \citep{2019ApJ...884...85C}. Thanks to deep imaging by {\it HST}, many multiple images have been identified for these clusters, leading to reliable measurements of dark matter distributions at the cores of 45 DM haloes. Measured ellipticities and orientations of dark matter distributions are then compared with those of Brightest Cluster Galaxies (BCGs). We also compare our results with theoretical model predictions based on the cosmological hydrodynamical simulation Horizon-AGN \citep{dubois14} to see if the observed ellipticities and alignments of orientations between BCGs and their host DM haloes are consistent with the $\Lambda$CDM model predictions.

The structure of this paper is as follows. In Section~\ref{sec:sample}, we describe the cluster sample and observational data used for the analysis. We present measurements of ellipticities and alignments in Section~\ref{sec:ell_pa_hst}, and the comparison with the Horizon-AGN simulation in Section~\ref{sec:sim_obs_ell}. We discuss results in Section~\ref{sec:discussions_comparison_obs}, and summarize results in Section~\ref{sec:summary_comparison_obs}. Throughout this paper, we adopt cosmological parameters based on the seven-year Wilkinson Microwave Anisotropy Probe \citep{WMAP}; the total matter density $\Omega_{{\rm m}} = 0.272$, cosmological constant $\Omega_{\Lambda} = 0.728$,  the baryon density $\Omega_{{\rm b}} = 0.045$, the amplitude of the power spectrum of density fluctuations $\sigma_{8} = 0.81$, the Hubble constant $H_0 = 70.4~{\rm km\,s^{-1}Mpc^{-1}}$, and the power-law index of the primordial power spectrum $n_{s} = 0.967$, which are the parameter set adopted in the Horizon-AGN simulation.

\section{Sample and data}\label{sec:sample}

\subsection{Cluster sample: HFF, CLASH, and RELICS} \label{sec:hst_sample}
We first describe how to measure ellipticities and orientations of galaxy clusters, all of which are observed with the {\it HST}. We use three survey data to construct our galaxy cluster sample: Hubble Frontier Field \footnote{https://archive.stsci.edu/prepds/frontier/} \citep[HFF;][]{2017ApJ...837...97L}, Cluster Lensing And Supernova survey with Hubble \footnote{https://archive.stsci.edu/prepds/clash/}  \citep[CLASH;][]{2012ApJS..199...25P}, and Reionization Lensing Cluster Survey \footnote{https://archive.stsci.edu/prepds/relics/} \citep[RELICS;][]{2019ApJ...884...85C}. Our strong lens mass models of 39 galaxy clusters from these three surveys (see Section~\ref{sec:method_sl}) are used to measure ellipticities and orientations of their dark matter distributions, whereas we use the {\it HST} images to measure light profiles of their BCGs (see Section~\ref{sec:method_bcg}).

\begin{table}
	\centering
	\caption{ Properties of our cluster sample. $M_{14}$ means virial mass of each cluster divided by $10^{14}M_{\odot}$. }
	\label{tab:clustersample}
	\begin{tabular}{ l l l l } 
		\hline
		survey & cluster name & $z$ & $M_{14}$ \\ 
\hline 
HFF & Abell~2744 & 0.308 & 18.0 \\  
HFF & MACS0416.1$-$2403 & 0.3971 & 12.0 \\  
HFF & MACS1149.5$+$2223 & 0.541 & 25.0 \\  
HFF & Abell~S1063 & 0.348 & 14.0 \\  
CLASH & Abell~209 & 0.206 & 11.5 \\  
CLASH & Abell~383 & 0.187 & 9.6 \\  
CLASH & MACS0329.7$-$0211 & 0.45 & 11.2 \\  
CLASH & MACS0429.6$-$0253 & 0.399 & 7.1 \\  
CLASH & MACS0744.9$+$3927 & 0.686 & 11.4 \\  
CLASH & Abell~611 & 0.288 & 12.4 \\  
CLASH & MACS1115.9$+$0129 & 0.355 & 12.0 \\  
CLASH & Abell~1423 & 0.213 & 10.9 \\  
CLASH & MACS1206.2$-$0847 & 0.439 & 18.9 \\  
CLASH & MACS1311.0$-$0310 & 0.494 & 6.5 \\  
CLASH & RXJ1347.5$-$1145 & 0.451 & 34.8 \\  
CLASH & MACS1720.3$+$3536 & 0.387 & 8.5 \\  
CLASH & Abell~2261 & 0.224 & 12.2 \\  
CLASH & MACS1931.8$-$2635 & 0.352 & 8.9 \\  
CLASH & RXJ2129.7$+$0005 & 0.234 & 7.6 \\  
CLASH & MS~2137$-$2353 & 0.313 & 7.4 \\  
CLASH & MACS0647.7$+$7015 & 0.584 & 24.3 \\  
CLASH & MACS2129.4$-$0741 & 0.57 & 12.6 \\  
RELICS & Abell~2163 & 0.203 & 28.3 \\  
RELICS & Abell~2537 & 0.2966 & 9.5 \\  
RELICS & Abell~3192 & 0.425 & 12.1 \\  
RELICS & Abell~697 & 0.282 & 19.0 \\  
RELICS & Abell~S295 & 0.3 & 11.7 \\  
RELICS & ACT-CL~J0102-49151 & 0.87 & 17.3 \\  
RELICS & CL~J0152.7-1357 & 0.833 & 11.3 \\  
RELICS & MACS~J0159.8-0849 & 0.405 & 12.2 \\  
RELICS & MACS~J0257.1-2325 & 0.5049 & 10.4 \\  
RELICS & MACS~J0308.9$+$2645 & 0.356 & 18.3 \\  
RELICS & MACSJ0417.5-1154 & 0.443 & 20.6 \\  
RELICS & MACS~J0553.4-3342 & 0.43 & 14.8 \\  
RELICS & PLCK~G171.9-40.7 & 0.27 & 18.5 \\  
RELICS & PLCK~G287.0$+$32.9 & 0.39 & 24.9 \\  
RELICS & RXC~J0142.9$+$4438 & 0.341 & 15.4 \\  
RELICS & RXC~J2211.7-0350 & 0.397 & 17.8 \\  
RELICS & SPT-CL~J0615-5746 & 0.972 & 10.8 \\  
		\hline
	\end{tabular}
\end{table}

\begin{figure*}
	\includegraphics[width=9cm]{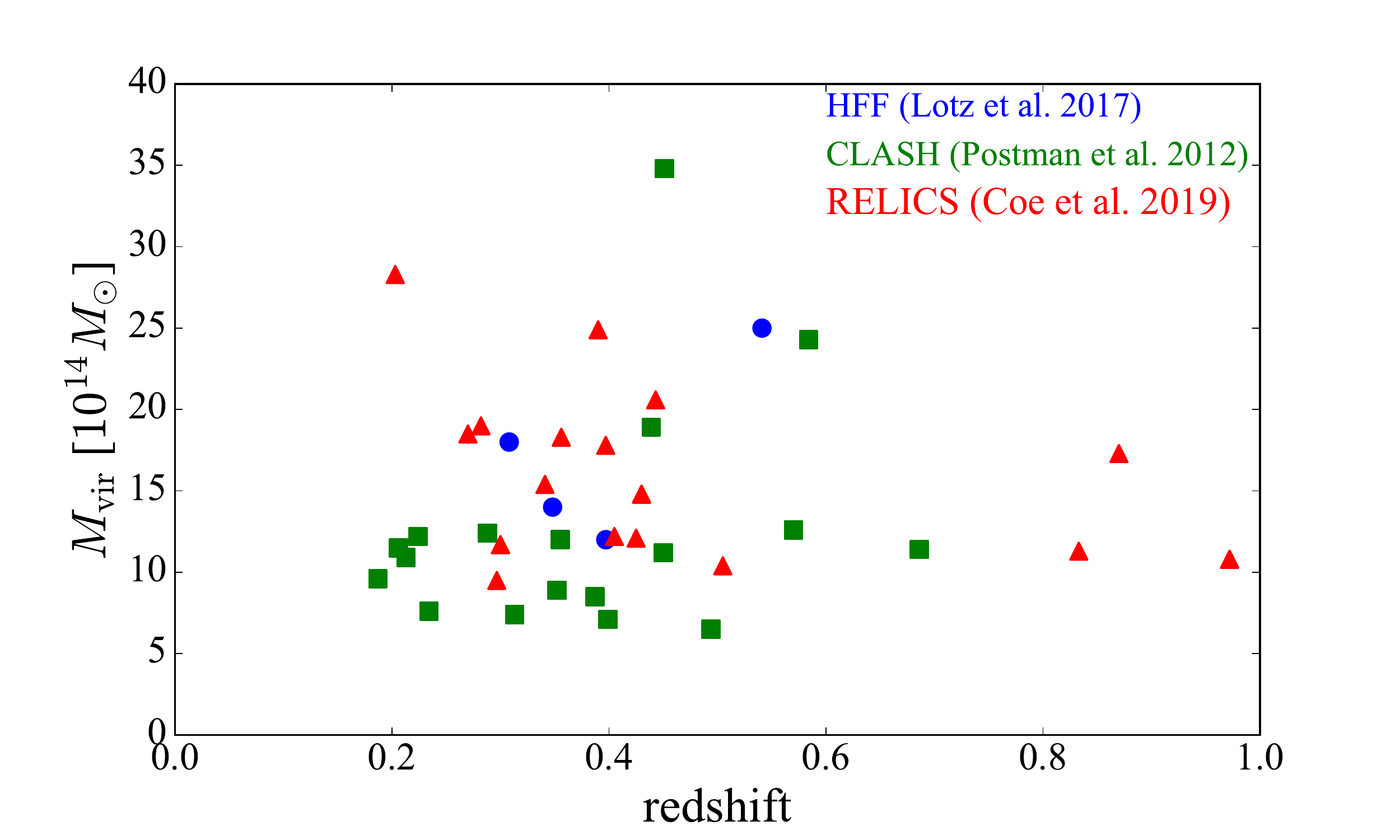}
    \caption{Virial masses and redshifts of clusters in our sample. Blue circles, green squares, and red triangles show clusters observed by HFF, CLASH, and RELICS, respectively.}
    \label{fig:z_mass_hst}
\end{figure*}

Table~\ref{tab:clustersample} summarizes properties of the galaxy cluster sample, with $M_{14}$ being the virial mass in units of $10^{14}M_{\odot}$. We compute their virial masses as follows. For HFF, we use $M_{\rm vir}$ shown in Table~2 of \citet{2017ApJ...837...97L}. For CLASH, first we convert the X-ray temperature shown in Table~4 of \citet{2012ApJS..199...25P} to $M_{500}$ by using an empirical relation \citep{2007A&A...474L..37A}. We then obtain $M_{\rm vir}$ from $M_{500}$ by assuming the \citet[][hereafter NFW]{NFW} profile with the concentration parameter of $c_{500} = 2.5$. For RELICS, we use $M_{500}$ shown in Table~2 of \citet{2019ApJ...884...85C} that are inferred from Sunyaev-Zel'dovich effect measurements by the Planck satellite and convert them to $M_{\rm vir}$  by assuming the NFW profile and $c_{500} = 2.5$. Figure~\ref{fig:z_mass_hst} summarizes virial masses and redshifts of galaxy clusters in our sample.

\subsection{Ellipticities and position angles of dark matter halo by strong lensing} \label{sec:method_sl}
We compare shapes of BCGs with those of dark matter distributions measured with strong lensing. See Appendix~\ref{sec:strong_lensing_method} for more details. In short, we use the software {\sc glafic} \citep{2010PASJ...62.1017O} for mass modeling, and reconstruct the mass distribution of each cluster assuming a parametric mass model that includes DM halo components modeled by an elliptical NFW profile as well as cluster member galaxy components (including BCG) modeled by an elliptical pseudo-Jaffe profile. More specifically, we introduce an ellipticity $e_{\rm SL}$ to the NFW profile simply by defining the convergence $\kappa$ as 
\begin{equation}
\kappa(x, y)=\kappa_{\rm NFW}\left(r=\sqrt{\frac{x^2}{1-e_{\rm SL}}+(1-e_{\rm SL})y^2}\right),
\end{equation}
where $\kappa_{\rm NFW}$ is the convergence profile of a spherical NFW profile \citep[e.g.,][]{1996A&A...313..697B} and $x$ and $y$ are coordinates along with minor and major axes of the ellipse. Therefore our definition of the ellipticity is $e=1-b/a$, where $a$ and $b$ are major and minor axis lengths of the ellipse. An additional model parameter for the elliptical NFW profile is the position angle $\theta_{\rm SL}$, which is defined as the polar angle of the major axis measured East of North. All model parameters including $e_{\rm SL}$ and $\theta_{\rm SL}$ of DM halo components are determined so as to reproduce observed multiple image positions (see also Appendix~\ref{sec:strong_lensing_method}). Since cluster member galaxies are modelled separately in our strong lens mass modeling, our measurements of halo shapes from the best-fitting $e_{\rm SL}$ and $\theta_{\rm SL}$ of DM halo components correspond to shapes of the smooth part of cluster DM haloes.

The precision and accuracy of strong lens mass modeling depend on the number of multiple images and the availability of spectroscopic redshifts for them. In order to obtain reliable measurements, we limit our analysis to clusters with three or more sets of multiple images. Since we are interested in comparing shapes of dark matter distributions with those of BCGs, we need to identify the corresponding BCG for each halo component. Specifically, we define it as the brightest cluster member galaxy among those located near the centre of a halo component.  We remove clusters if identifications of BCGs are not secure due to large offsets ($\ga 5''$) between halo components and putative BCGs or no obvious bright galaxies near halo centres. Such situation can be seen in complex merging clusters such as MACSJ0717.5+3745 in HFF. Clusters listed in Table~\ref{tab:clustersample} and Appendix~\ref{sec:strong_lensing_method} are those after these selections of the number of multiple images and the secure BCG identification are applied.

In some of the 39 clusters in our cluster sample, there are more than one prominent halo components. If their model parameters are well constrained by strong lensing data and bright central galaxies are securely identified for them, we include multiple halo components from a single cluster separately in our analysis. Since 6 out of the 39 clusters have two separate ``haloes'', we measure shapes of 45 DM haloes in total.

\subsection{Ellipticities and position angles of BCGs} \label{sec:method_bcg}
We measure shapes of 45 BCGs at the centres of DM haloes whose shapes are measured by strong lensing. For all the BCGs, we use {\it HST} images in F814W band and calculate ellipticities and position angles using a tensor method following \citet{okabe18} and \citet{2019arXiv191104653O}. Specifically, we compute the mass tensor
\begin{equation}
I_{{\rm BCG},\alpha\beta}=\sum_{i,j}\Sigma(i,j)\left[x_\alpha(i,j)-x_{\alpha}^{\rm CSB}\right]\left[x_\beta(i,j)-x_{\beta}^{\rm CSB}\right],
\end{equation}
where $\alpha$, $\beta=1$, $2$ labels the two-dimensional coordinates of the image, $\Sigma(i,j)$ denotes the F814W-band surface brightness of the BCG at the pixel ($i$, $j$), and $x_\alpha(i,j)-x_{\alpha}^{\rm CSB}$ is the projected position relative to the centre of the surface brightness defined as 
\begin{equation}
x_{\alpha}^{\rm CSB} \equiv \frac{\sum_{i,j}\Sigma(i,j) x_\alpha(i,j)}{\sum_{i,j}\Sigma(i,j)}.
\end{equation}
We start with an ellipse fit within a circle of a given radius, diagonalize the tensor to obtain the axis ratio $b/a$ and the position angle. We repeat the ellipse fit within an ellipse with the axis ratio $b/a$ from the previous fit. Fitting is repeated until the fractional difference of eigenvalues of the tensor between the previous and new fits becomes smaller than $10^{-8}$, and $e_{\rm BCG}=1-b/a$ is adopted as the ellipticity of the BCG. From the converged mass tensor we also derive the position angle $\theta_{\rm BCG}$ measured East of North.
Since the surface brightness distribution of galaxies in the red wavelength reflects the stellar mass distribution reasonable well, we assume that the measured ellipticity and position angle corresponds to those of the stellar mass distribution in the BCG when comparing our results with simulations.
We consider three radii, $R_{ab}\equiv\sqrt{ab} = 10$, 20, and 30~pkpc, where pkpc denotes kpc in physical (instead of comoving) scales, for the radius of the ellipse for fitting. The choice of the value $\sqrt{ab}$ is somewhat arbitrary but roughly corresponds to typical scales of BCGs. We measure ellipticities at the three scales in order to discuss the effect of satellite galaxies around the BCGs because we do not remove them in the ellipse fit procedure. Then we adopt 20~pkpc as a fiducial scale throughout the paper. We note that the average size of point spread function (PSF) of the {\it HST} images of $\sim97$ milliarcseconds \citep[e.g.,][]{2007ApJS..172...38S, 2007ApJS..172..196K}, corresponding to a physical scale of $\sim500$~pc at the mean redshift of clusters $\langle z \rangle \sim0.4$, is much smaller than the ellipse scales,  and thus the effect of PSF can be safely ignored in the ellipse fit procedure.

\begin{figure*}
   \includegraphics[width=4.2cm]{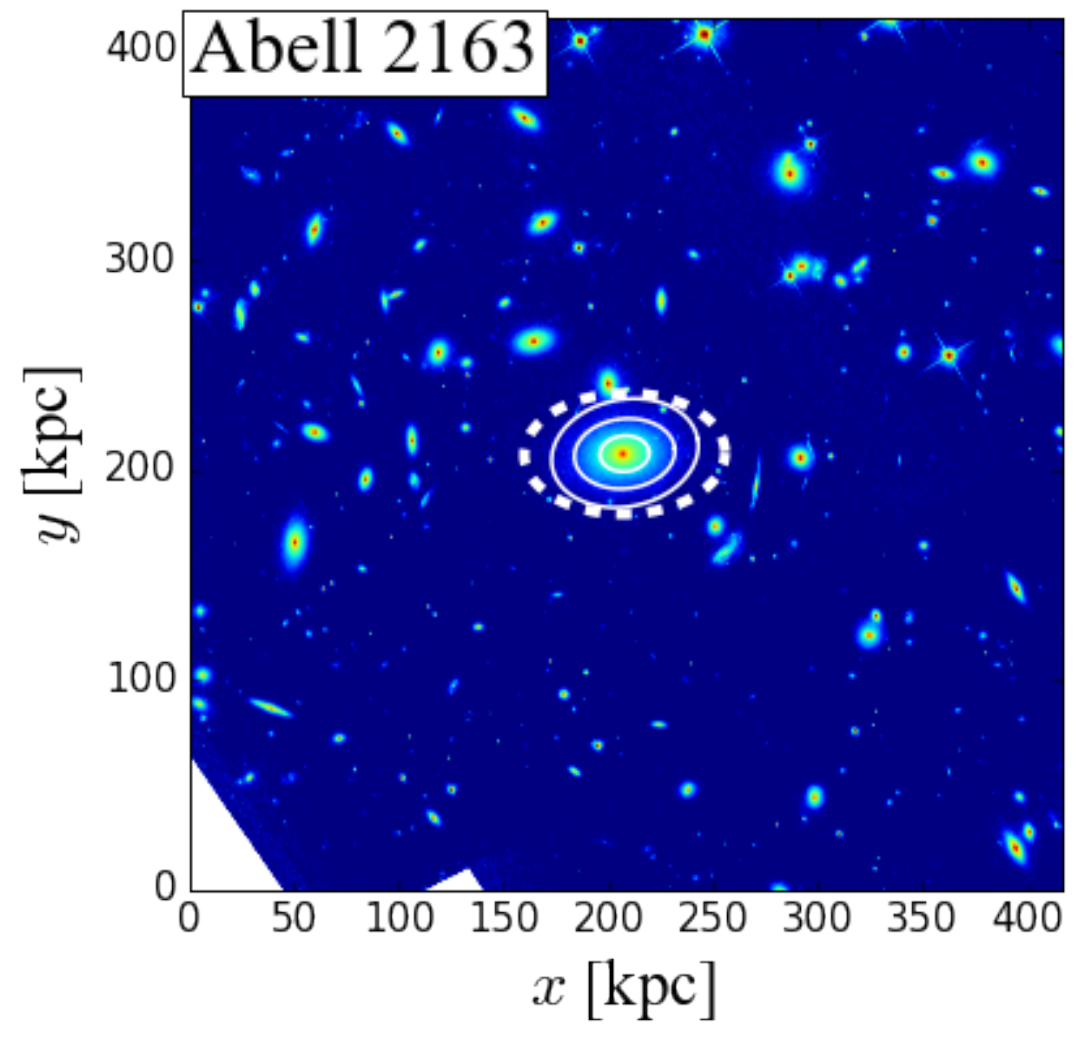}
   \includegraphics[width=4.2cm]{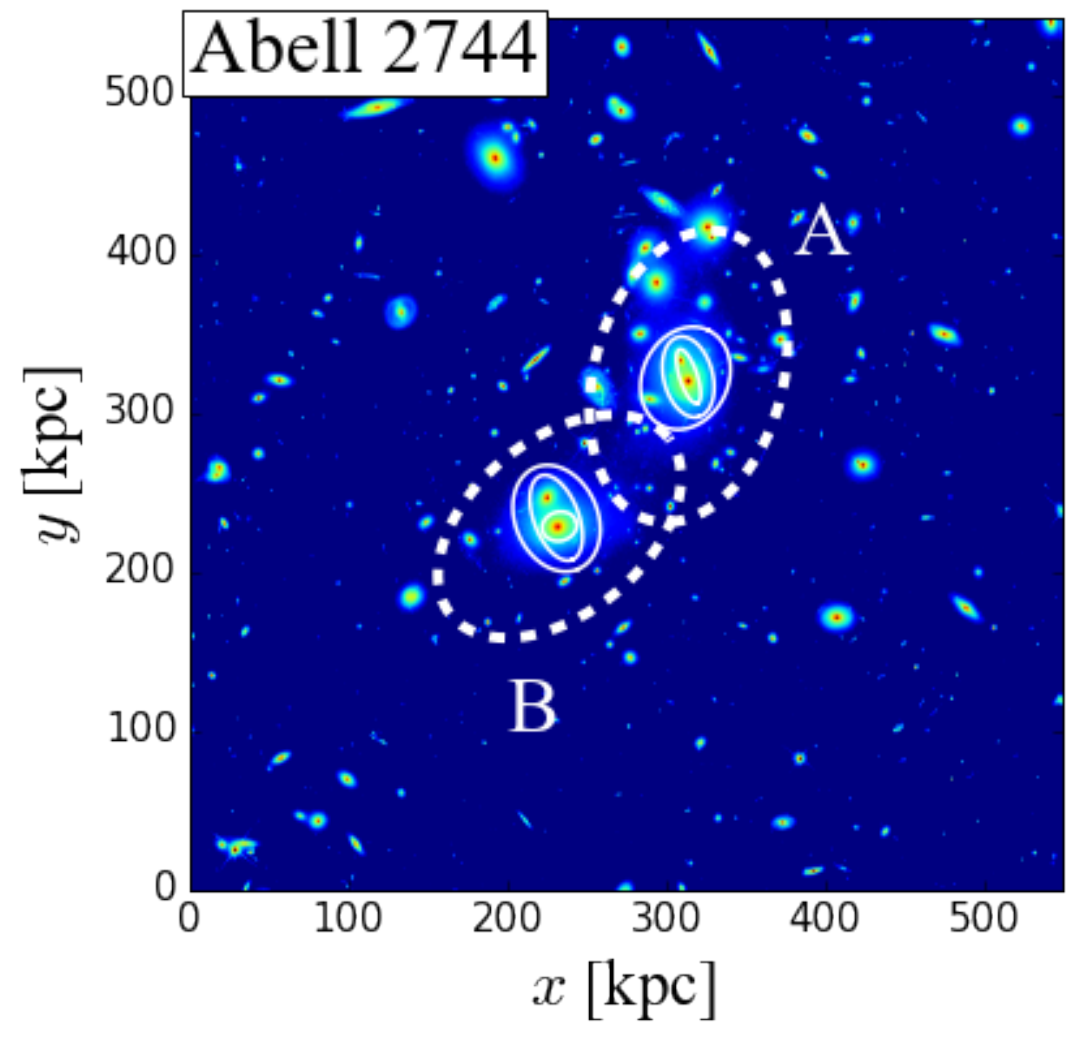}
  \caption{{\it Left:} A F814W band {\it HST} image of Abell 2163. Solid lines correspond to fitted ellipses of BCGs at $R_{ab} = 10$, 20, and 30~pkpc. The scale of DM halo ({\it dashed line}) is set to the Einstein radius with source redshift $z_{s} = 3.0$. {\it Right:} Similar to the left panel, but for the double-peak cluster Abell 2744. }
   \label{fig:abell_images}
\end{figure*}

Figure~\ref{fig:abell_images} shows examples of F814W band {\it HST} images of single-peak ({\it left}) and double-peak ({\it right}) clusters. The ellipse scales of DM haloes in the figure are the Einstein radii for a typical source redshift $z_{s} = 3.0$ probed by strong lensing method. As mentioned above, satellite galaxies in clusters indeed affect ellipse fits of BCGs such that their effects tend to be more significant for larger $R_{ab}$. Table~\ref{tab:bcg_sample_1} summarizes derived ellipticities and position angles of DM haloes and BCGs.

\begin{table*}
	\centering
	\caption{Properties of BCGs and measured value of ellipticities and position angles of BCGs and their host DM haloes. The symbols $e_{\rm BCG}^{R_{ab}}$ and $\theta_{\rm BCG}^{R_{ab}}$ denote the ellipticities and position angles of BCGs at scale of $R_{ab}$, and $e_{\rm SL}$ and $\theta_{\rm SL}$ are those of DM haloes. The position angles are in degree measured East of North.
	}
	\label{tab:bcg_sample_1}
	{\footnotesize
	\begin{tabular}{ l l l l l l l l l l l l } 
		\hline
survey & BCG name & ra & dec & $e_{\rm BCG}^{10}$ & $\theta_{\rm BCG}^{10}$ & $e_{\rm BCG}^{20}$ & $\theta_{\rm BCG}^{20}$ & $e_{\rm BCG}^{30}$ & $\theta_{\rm BCG}^{30}$ & $e_{\rm SL}$ & $\theta_{\rm SL}$ \\ 
\hline 
HFF & Abell~2744~A & $3.5862553$ & $-30.4001723$ & $0.686$ & $18.1$ & $0.404$ & $14.52$ & $0.193$ & $-25.93$ & $0.365^{+0.031}_{-0.028}$ & $-14.95^{+4.04}_{-2.88}$ \\[2pt]  
HFF & Abell~2744~B & $3.5920369$ & $-30.405741$ & $0.165$ & $-76.09$ & $0.498$ & $20.54$ & $0.252$ & $23.31$ & $0.379^{+0.021}_{-0.024}$ & $-50.58^{+1.79}_{-2.02}$ \\[2pt]  
HFF & MACS0416.1$-$2403~A & $64.0380978$ & $-24.0674837$ & $0.214$ & $55.66$ & $0.339$ & $52.15$ & $0.409$ & $37.03$ & $0.661^{+0.009}_{-0.01}$ & $60.58^{+0.65}_{-1.07}$ \\[2pt]  
HFF & MACS0416.1$-$2403~B & $64.0436968$ & $-24.0729844$ & $0.214$ & $76.19$ & $0.494$ & $40.31$ & $0.415$ & $40.61$ & $0.693^{+0.017}_{-0.017}$ & $42.92^{+0.97}_{-1.1}$ \\[2pt]  
HFF & MACS1149.5$+$2223 & $177.3987502$ & $22.3985322$ & $0.256$ & $7.94$ & $0.303$ & $-52.27$ & $0.657$ & $-42.83$ & $0.493^{+0.021}_{-0.018}$ & $-53.63^{+1.3}_{-1.28}$ \\[2pt]  
HFF & Abell~S1063 & $342.1832095$ & $-44.5308829$ & $0.204$ & $-13.69$ & $0.297$ & $30.9$ & $0.27$ & $47.1$ & $0.454^{+0.011}_{-0.011}$ & $53.38^{+0.33}_{-0.35}$ \\[2pt]  
CLASH & Abell~209 & $22.9689565$ & $-13.6112333$ & $0.203$ & $-43.94$ & $0.361$ & $-28.08$ & $0.227$ & $-38.59$ & $0.71^{+0.063}_{-0.159}$ & $309.5^{+3.69}_{-4.11}$ \\[2pt]  
CLASH & Abell~383 & $42.0140947$ & $-3.5292113$ & $0.128$ & $13.43$ & $0.105$ & $16.51$ & $0.319$ & $74.11$ & $0.216^{+0.059}_{-0.039}$ & $13.8^{+2.42}_{-3.78}$ \\[2pt]  
CLASH & MACS0329.7$-$0211 & $52.4232222$ & $-2.1962171$ & $0.175$ & $-49.94$ & $0.184$ & $-28.5$ & $0.226$ & $-37.02$ & $0.25^{+0.051}_{-0.044}$ & $-17.35^{+10.59}_{-6.71}$ \\[2pt]  
CLASH & MACS0429.6$-$0253 & $67.4000333$ & $-2.8851685$ & $0.274$ & $8.55$ & $0.316$ & $-5.0$ & $0.367$ & $0.92$ & $0.462^{+0.055}_{-0.062}$ & $-9.72^{+0.64}_{-0.62}$ \\[2pt]  
CLASH & MACS0744.9$+$3927 & $116.2199938$ & $39.4574046$ & $0.161$ & $15.79$ & $0.331$ & $21.01$ & $0.274$ & $19.14$ & $0.073^{+0.063}_{-0.042}$ & $-47.64^{+18.88}_{-18.75}$ \\[2pt]  
CLASH & Abell~611 & $120.2367241$ & $36.0565643$ & $0.174$ & $40.54$ & $0.291$ & $31.97$ & $0.56$ & $21.21$ & $0.257^{+0.025}_{-0.023}$ & $41.06^{+0.92}_{-1.08}$ \\[2pt]  
CLASH & MACS1115.9$+$0129 & $168.9662572$ & $1.4986333$ & $0.256$ & $-35.92$ & $0.381$ & $-35.21$ & $0.318$ & $-39.44$ & $0.607^{+0.066}_{-0.073}$ & $142.51^{+1.17}_{-1.32}$ \\[2pt]  
CLASH & Abell~1423 & $179.322349$ & $33.6109896$ & $0.288$ & $59.48$ & $0.364$ & $59.53$ & $0.424$ & $59.16$ & $0.233^{+0.206}_{-0.114}$ & $47.17^{+23.91}_{-25.98}$ \\[2pt]  
CLASH & MACS1206.2$-$0847 & $181.5506031$ & $-8.80093$ & $0.422$ & $-75.41$ & $0.523$ & $-75.07$ & $0.294$ & $88.52$ & $0.536^{+0.018}_{-0.017}$ & $109.15^{+0.61}_{-0.56}$ \\[2pt]  
CLASH & MACS1311.0$-$0310 & $197.7575102$ & $-3.1777062$ & $0.133$ & $-43.7$ & $0.2$ & $-49.03$ & $0.347$ & $-69.66$ & $0.421^{+0.076}_{-0.075}$ & $10.92^{+2.04}_{-2.22}$ \\[2pt]  
CLASH & RXJ1347.5$-$1145~A & $206.8775419$ & $-11.7526347$ & $0.17$ & $-4.98$ & $0.149$ & $-14.96$ & $0.239$ & $-3.28$ & $0.432^{+0.048}_{-0.042}$ & $13.79^{+5.78}_{-4.59}$ \\[2pt]  
CLASH & RXJ1347.5$-$1145~B & $206.8825922$ & $-11.7531986$ & $0.376$ & $33.57$ & $0.358$ & $30.0$ & $0.429$ & $16.06$ & $0.696^{+0.071}_{-0.115}$ & $29.01^{+2.07}_{-1.97}$ \\[2pt]  
CLASH & MACS1720.3$+$3536 & $260.0697955$ & $35.6073118$ & $0.19$ & $-1.31$ & $0.191$ & $-9.27$ & $0.396$ & $-2.12$ & $0.272^{+0.042}_{-0.052}$ & $5.2^{+2.24}_{-4.36}$ \\[2pt]  
CLASH & Abell~2261 & $260.6130615$ & $32.1326534$ & $0.061$ & $-18.87$ & $0.147$ & $-6.03$ & $0.112$ & $-7.0$ & $0.2^{+0.027}_{-0.028}$ & $46.36^{+2.33}_{-3.46}$ \\[2pt]  
CLASH & MACS1931.8$-$2635 & $292.9567874$ & $-26.575729$ & $0.162$ & $-24.71$ & $0.467$ & $-25.8$ & $0.448$ & $-26.85$ & $0.459^{+0.017}_{-0.022}$ & $-5.03^{+1.03}_{-0.88}$ \\[2pt]  
CLASH & RXJ2129.7$+$0005 & $322.4164769$ & $0.0892336$ & $0.405$ & $68.4$ & $0.476$ & $55.15$ & $0.454$ & $66.89$ & $0.547^{+0.045}_{-0.036}$ & $67.51^{+0.88}_{-0.74}$ \\[2pt] 
CLASH & MS~2137$-$2353 & $325.0631662$ & $-23.6611459$ & $0.099$ & $62.61$ & $0.183$ & $-30.62$ & $0.068$ & $-84.57$ & $0.204^{+0.055}_{-0.05}$ & $60.52^{+3.43}_{-2.36}$ \\[2pt]  
CLASH & MACS0647.7$+$7015 & $101.9610124$ & $70.2483297$ & $0.414$ & $-71.06$ & $0.778$ & $-71.88$ & $0.694$ & $-72.56$ & $0.787^{+0.009}_{-0.018}$ & $104.91^{+0.41}_{-0.45}$ \\[2pt]  
CLASH & MACS2129.4$-$0741 & $322.3587881$ & $-7.6910536$ & $0.401$ & $80.12$ & $0.759$ & $76.21$ & $0.663$ & $79.4$ & $0.576^{+0.041}_{-0.047}$ & $81.2^{+1.34}_{-1.24}$ \\[2pt]  
RELICS & Abell~2163 & $243.9539405$ & $-6.1448406$ & $0.271$ & $-85.83$ & $0.307$ & $-83.8$ & $0.278$ & $-78.32$ & $0.398^{+0.057}_{-0.061}$ & $91.04^{+1.92}_{-2.09}$ \\[2pt]  
RELICS & Abell~2537 & $347.0925316$ & $-2.1920915$ & $0.235$ & $-53.04$ & $0.51$ & $-53.46$ & $0.433$ & $-55.26$ & $0.391^{+0.049}_{-0.048}$ & $-57.48^{+2.03}_{-1.18}$ \\[2pt]  
RELICS & Abell~3192 & $59.7253299$ & $-29.9252985$ & $0.654$ & $61.49$ & $0.557$ & $59.16$ & $0.572$ & $54.43$ & $0.557^{+0.112}_{-0.096}$ & $71.87^{+7.56}_{-9.16}$ \\[2pt]  
RELICS & Abell~697 & $130.7398208$ & $36.3664976$ & $0.513$ & $22.0$ & $0.277$ & $13.39$ & $0.236$ & $-4.84$ & $0.516^{+0.153}_{-0.132}$ & $-25.61^{+2.71}_{-1.91}$ \\[2pt]  
RELICS & Abell~S295~A & $41.3533874$ & $-53.0293239$ & $0.244$ & $-54.89$ & $0.369$ & $-57.57$ & $0.671$ & $73.81$ & $0.668^{+0.076}_{-0.11}$ & $-51.58^{+2.99}_{-3.05}$ \\[2pt]  
RELICS & Abell~S295~B & $41.3956943$ & $-53.048456$ & $0.021$ & $84.56$ & $0.641$ & $20.31$ & $0.488$ & $18.58$ & $0.732^{+0.045}_{-0.079}$ & $-27.57^{+1.97}_{-1.86}$ \\[2pt]  
RELICS & ACT-CL~J0102-49151 & $15.7406954$ & $-49.2720008$ & $0.479$ & $-48.66$ & $0.429$ & $-46.5$ & $0.493$ & $-45.9$ & $0.637^{+0.052}_{-0.039}$ & $-61.95^{+2.62}_{-4.55}$ \\[2pt]  
RELICS & CL~J0152.7-1357 & $28.1824343$ & $-13.955155$ & $0.66$ & $-60.18$ & $0.372$ & $-68.61$ & $0.43$ & $-69.49$ & $0.683^{+0.07}_{-0.098}$ & $37.2^{+2.08}_{-4.79}$ \\[2pt]  
RELICS & MACS~J0159.8-0849 & $29.9554505$ & $-8.8329993$ & $0.45$ & $-73.03$ & $0.133$ & $-46.31$ & $0.151$ & $88.02$ & $0.345^{+0.108}_{-0.085}$ & $66.19^{+3.77}_{-4.38}$ \\[2pt]  
RELICS & MACS~J0257.1-2325 & $44.2864412$ & $-23.4346896$ & $0.273$ & $86.84$ & $0.383$ & $82.99$ & $0.456$ & $81.37$ & $0.773^{+0.017}_{-0.023}$ & $88.23^{+1.08}_{-1.14}$ \\[2pt]  
RELICS & MACS~J0308.9$+$2645 & $47.2331706$ & $26.760531$ & $0.676$ & $-16.35$ & $0.454$ & $65.47$ & $0.323$ & $63.51$ & $0.191^{+0.03}_{-0.036}$ & $60.29^{+0.94}_{-0.9}$ \\[2pt]  
RELICS & MACSJ0417.5-1154 & $64.3945535$ & $-11.9088405$ & $0.533$ & $-31.88$ & $0.462$ & $-31.04$ & $0.491$ & $-12.17$ & $0.667^{+0.02}_{-0.03}$ & $-33.87^{+0.56}_{-0.5}$ \\[2pt]  
RELICS & MACS~J0553.4-3342~A & $88.357296$ & $-33.7076965$ & $0.175$ & $-85.07$ & $0.292$ & $-83.51$ & $0.652$ & $-83.0$ & $0.696^{+0.04}_{-0.055}$ & $93.58^{+0.7}_{-0.6}$ \\[2pt]  
RELICS & MACS~J0553.4-3342~B & $88.3306883$ & $-33.7075393$ & $0.747$ & $-28.01$ & $0.609$ & $-40.65$ & $0.508$ & $-54.07$ & $0.414^{+0.22}_{-0.158}$ & $111.16^{+17.06}_{-13.3}$ \\[2pt]  
RELICS & PLCK~G171.9-40.7 & $48.2394369$ & $8.369767$ & $0.133$ & $-55.4$ & $0.293$ & $-54.83$ & $0.679$ & $-51.3$ & $0.692^{+0.031}_{-0.031}$ & $-35.91^{+1.32}_{-1.72}$ \\[2pt]  
RELICS & PLCK~G287.0$+$32.9 & $177.7089998$ & $-28.0821435$ & $0.185$ & $-29.57$ & $0.282$ & $-36.87$ & $0.693$ & $-65.23$ & $0.577^{+0.054}_{-0.078}$ & $-35.51^{+2.96}_{-1.59}$ \\[2pt]  
RELICS & RXC~J0142.9$+$4438 & $25.7300898$ & $44.6346655$ & $0.388$ & $-25.64$ & $0.342$ & $-24.02$ & $0.507$ & $-47.01$ & $0.233^{+0.025}_{-0.027}$ & $-19.54^{+0.46}_{-0.45}$ \\[2pt]  
RELICS & RXC~J2211.7-0350 & $332.9413416$ & $-3.8289814$ & $0.224$ & $8.36$ & $0.273$ & $12.72$ & $0.321$ & $17.8$ & $0.469^{+0.047}_{-0.049}$ & $8.41^{+2.07}_{-1.05}$ \\[2pt]  
RELICS & SPT-CL~J0615-5746~A & $93.9654777$ & $-57.7801148$ & $0.388$ & $27.37$ & $0.67$ & $18.65$ & $0.703$ & $23.67$ & $0.539^{+0.091}_{-0.104}$ & $14.24^{+8.22}_{-7.75}$ \\[2pt]  
RELICS & SPT-CL~J0615-5746~B & $93.9703845$ & $-57.7753024$ & $0.368$ & $81.72$ & $0.119$ & $-64.51$ & $0.798$ & $84.73$ & $0.524^{+0.082}_{-0.116}$ & $25.55^{+4.84}_{-4.55}$ \\[2pt]  
		\hline
	\end{tabular}
        }
\end{table*}

\section{Ellipticities and alignment angles between dark matter haloes and BCGs in the {\it HST} cluster sample} \label{sec:ell_pa_hst}
\begin{figure*}
\centering
	\includegraphics[width=8.5cm]{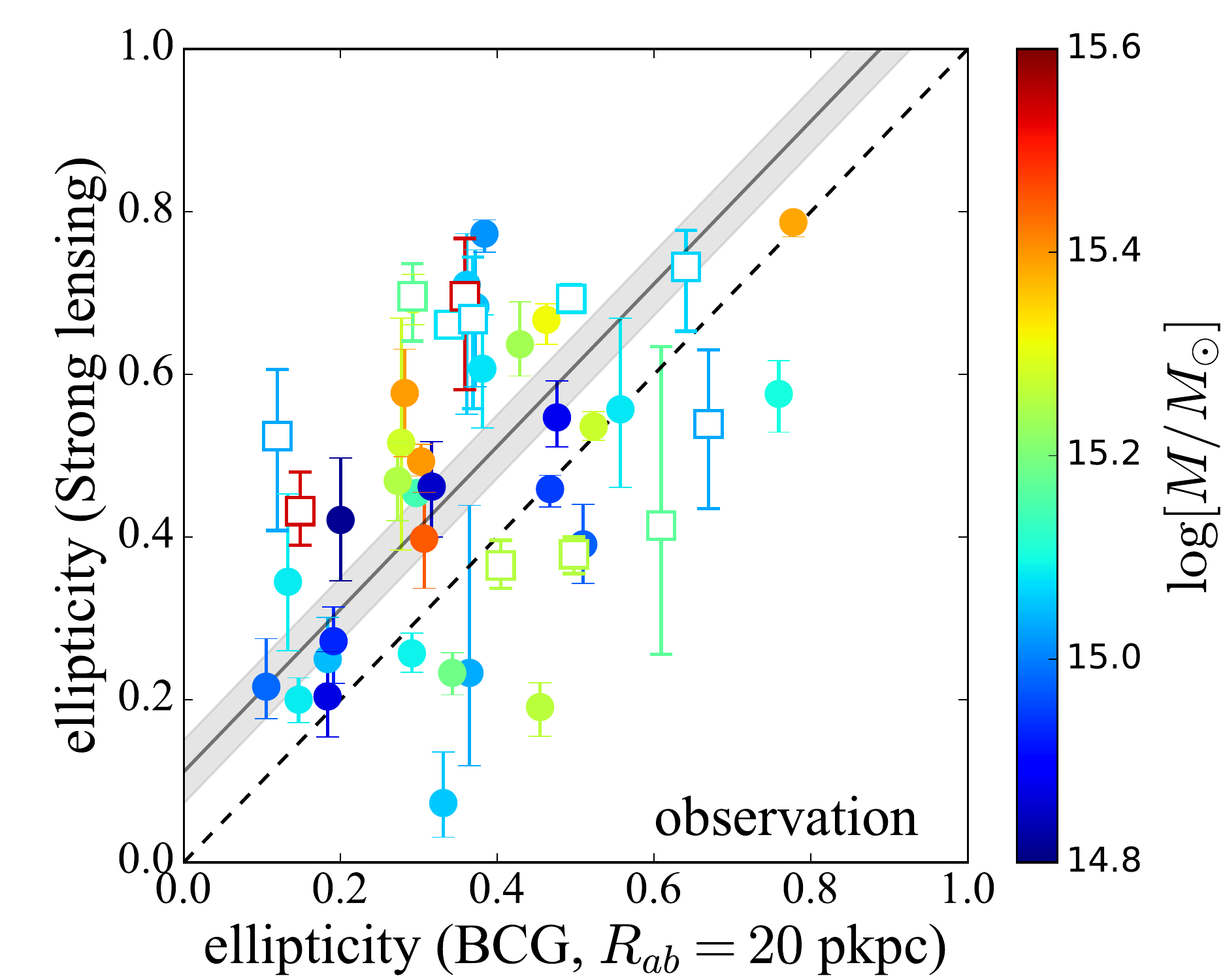}
    \caption{
    Ellipticities of DM haloes against those of BCGs fitted at the fiducial scale $R_{ab} = 20$~pkpc. Colour corresponds to the cluster mass (see Table~\ref{tab:clustersample}). Filled circles and open squares indicate single- and double-peak clusters, respectively. The dashed line indicates that the ellipticity of DM haloes and BCGs are the same.  The solid line with shading region shows the mean values and its error of differences between $e_{\rm SL}$ and $e_{\rm BCG}^{20}$. }
    \label{fig:ell_sl_bcg_hst}
\end{figure*}

Figure~\ref{fig:ell_sl_bcg_hst} shows the correlation between ellipticities of DM haloes from strong lensing method and those of BCGs from the {\it HST} images. The DM haloes are on average more elliptical than their BCGs. Specifically, the mean value of differences of ellipticities is $\langle e_{\rm SL} - e_{\rm BCG}\rangle = 0.11\pm0.03$, where $e_{\rm SL}$  and $e_{\rm BCG}$ denote ellipticities of DM haloes measured by strong lensing  and those of BCGs at the fiducial scale $R_{ab}=20$~pkpc, respectively.\footnote{When we assume a more general form, $e_{\rm SL}=ae_{\rm BCG}+b$, we obtain $a=0.518_{-0.157}^{+0.162}$ and $b=0.290_{-0.065}^{+0.064}$.} This result appears to be inconsistent with \citet{okabe18}, who find that ellipticities of DM distribution and those of central galaxies of cluster-sized haloes are similar in the Horizon-AGN simulation despite with large scatters. We will make more careful comparison with the Horizon-AGN simulation in Section~\ref{sec:sim_obs_ell}. Figure~\ref{fig:ell_sl_bcg_hst} also indicates that the correlation between $e_{\rm SL}$ and $e_{\rm BCG}$ is not tight. This result is in line with \citet{2008MNRAS.390.1562H}, who showed that ellipticities of BCGs and X-ray surface brightness of their host clusters are not strongly correlated. 

\begin{figure*}
\centering
	\includegraphics[width=8.5cm]{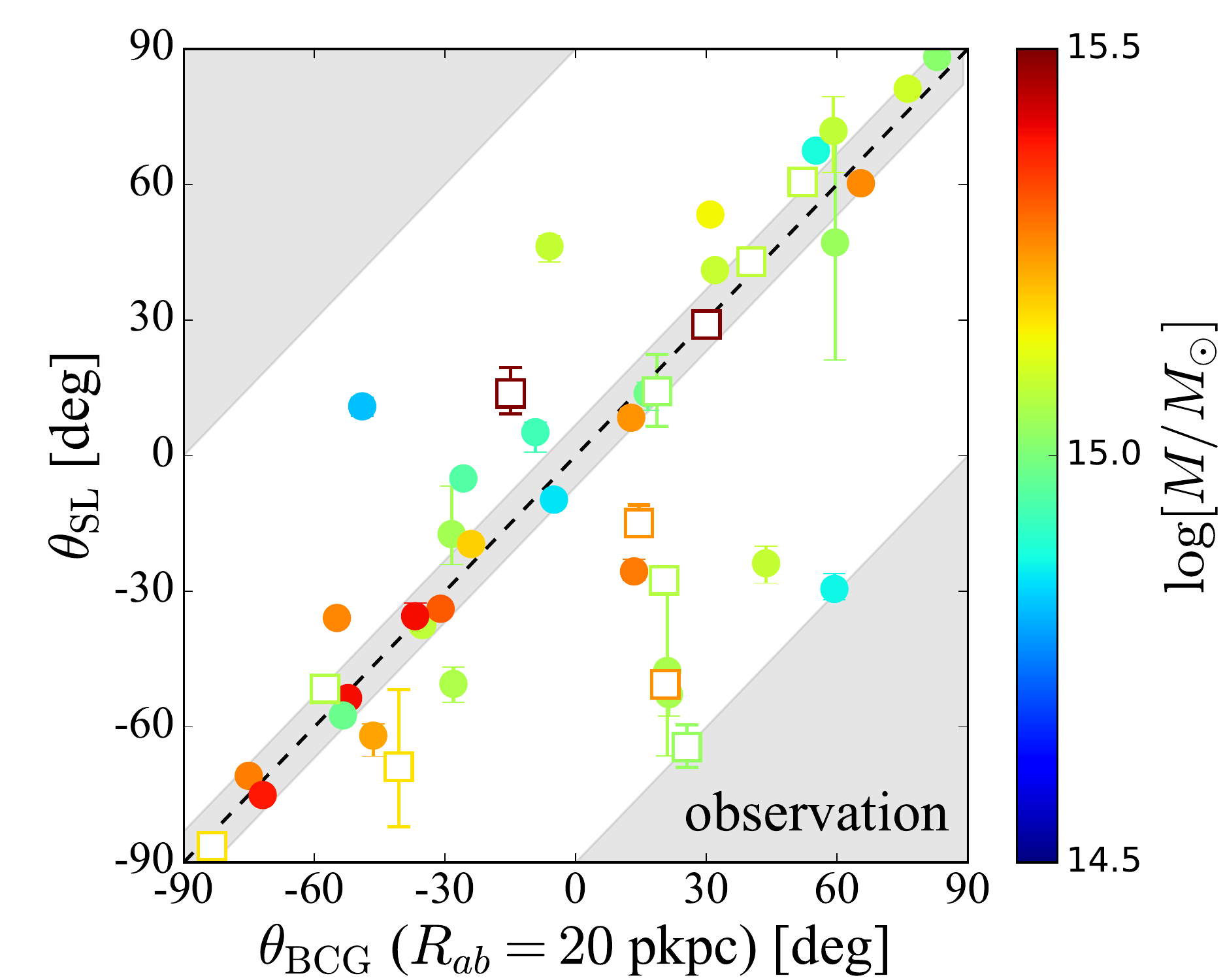}
    \caption{
    Position angles of DM haloes against those of BCGs fitted at the fiducial scale $R_{ab} = 20$~pkpc. Symbols are same as in Figure~\ref{fig:ell_sl_bcg_hst}.
    The dashed line indicates the case of the perfect alignment, $\theta_{\rm SL} = \theta^{20}_{\rm BCG}$. The shaded region around the dashed line shows the error of differences between $\theta_{\rm SL}$ and $\theta_{\rm BCG}^{20}$.
    The shaded regions at right bottom and top left have misalignment angles larger than $90^{\circ}$,
    $|\theta_{\rm SL} - \theta_{\rm BCG}^{20}| > 90^{\circ}$, and thus position angles of clusters in this regions are shifted by $90^{\circ}$ to locate them in the proper position.
    }
    \label{fig:theta_sl_bcg_hst}
\end{figure*}

Figure~\ref{fig:theta_sl_bcg_hst} plots the correlation between position angles of DM haloes and BCGs, indicating that they are well aligned. 
This result is consistent with recent observational results by \citet{2019arXiv191106333H} which shows that the BCGs and DM haloes are well aligned with each other, and also is qualitatively consistent with those of cluster-sized haloes in the Horizon-AGN simulation shown in \citet{okabe18}. We present further comparison with the Horizon-AGN simulation in Section~\ref{sec:sim_obs_ell}.

\begin{figure*}
\centering
	\includegraphics[width=8.5cm]{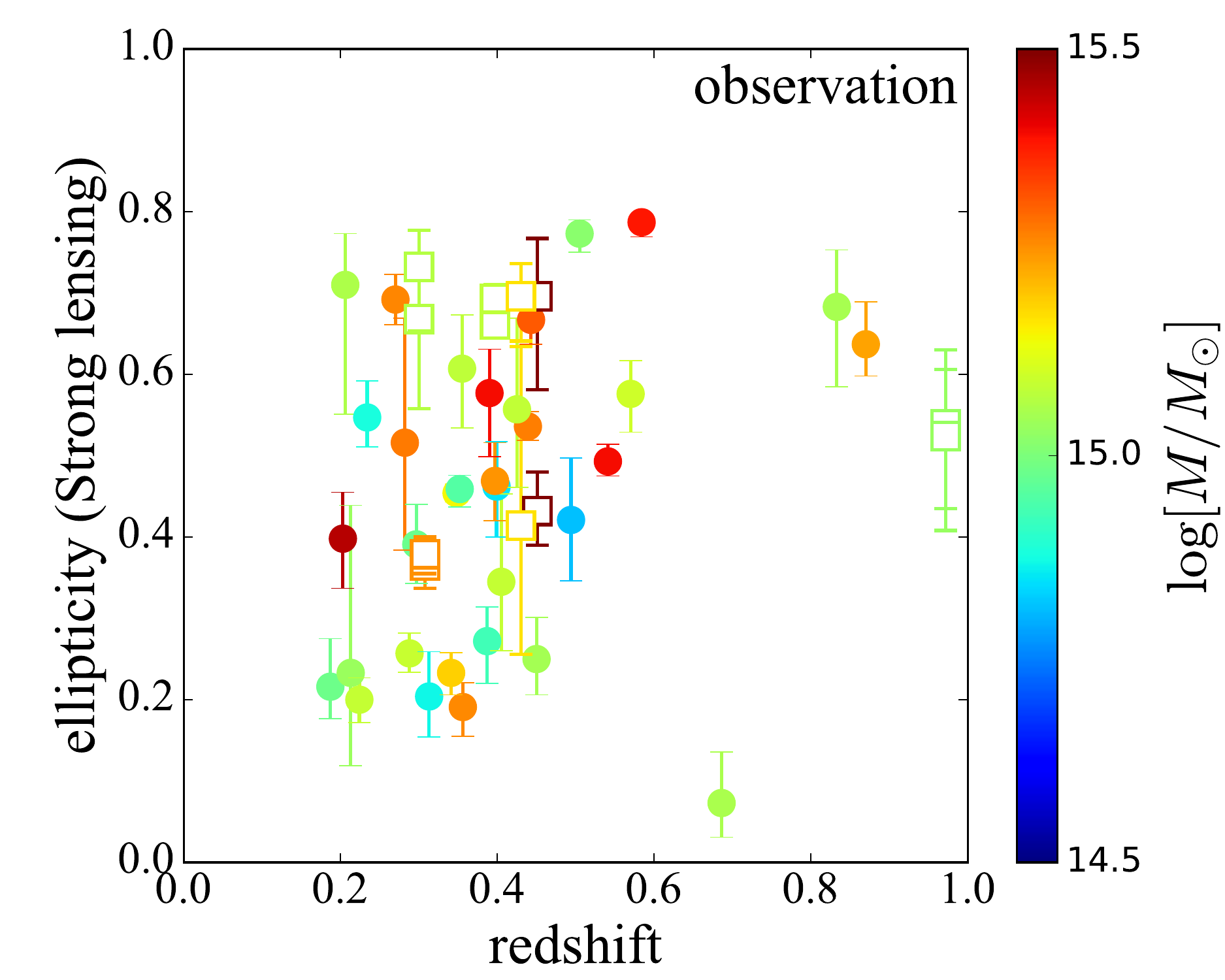}
    \caption{
    Ellipticities of DM haloes measured by strong lensing as a function of redshift. Symbols are same as in Figure~\ref{fig:ell_sl_bcg_hst}.
        }
    \label{fig:ell_sl_z_hst}
\end{figure*}

\begin{figure*}
\centering
	\includegraphics[width=8.5cm]{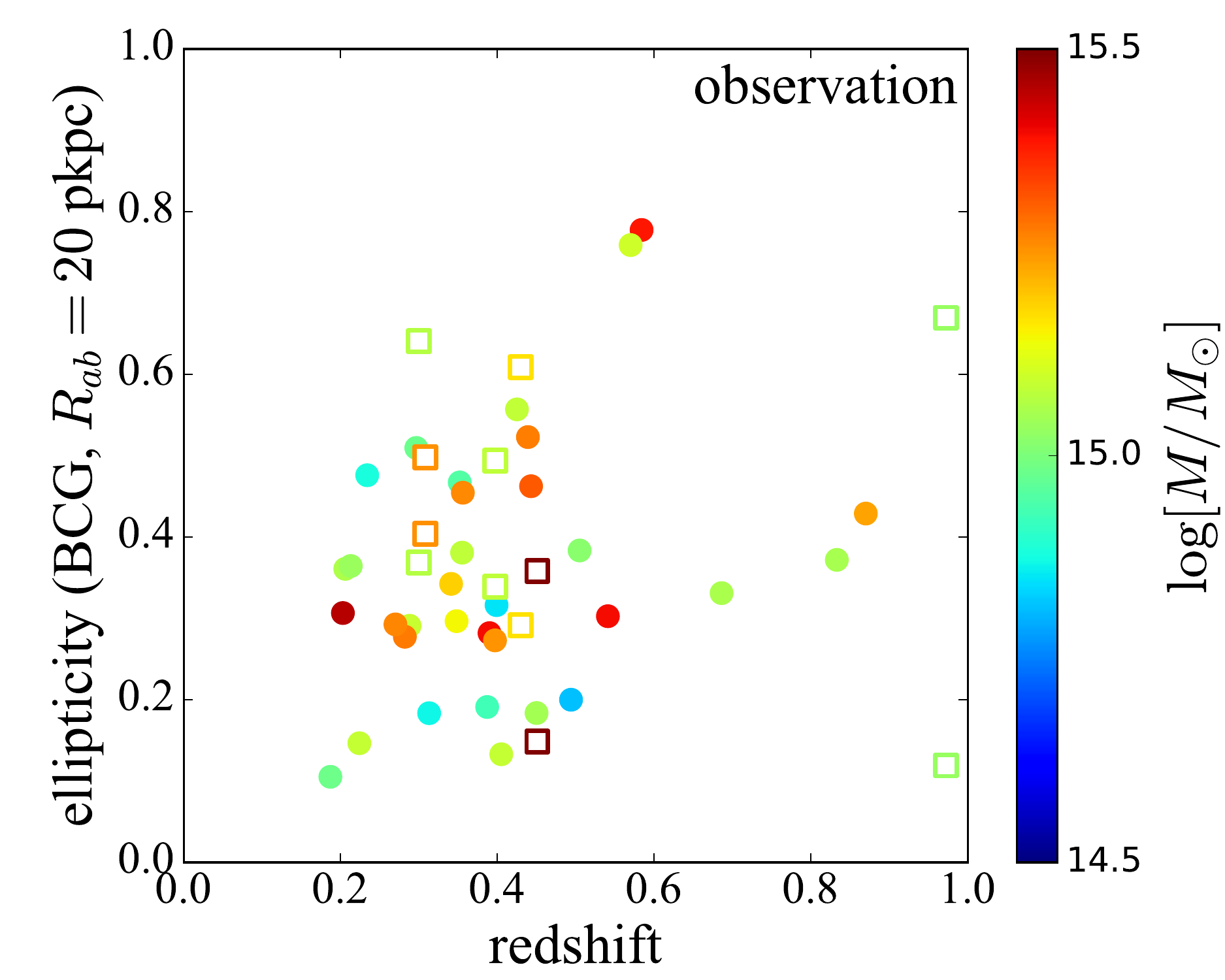}
    \caption{
    Ellipticities of BCGs fitted at the fiducial scale $R_{ab} = 20$~pkpc for as a function of redshift. Symbols are same as in Figure~\ref{fig:ell_sl_bcg_hst}.
   }
    \label{fig:ell_bcg_z_hst}
\end{figure*}

\begin{figure*}
\centering
	\includegraphics[width=8.5cm]{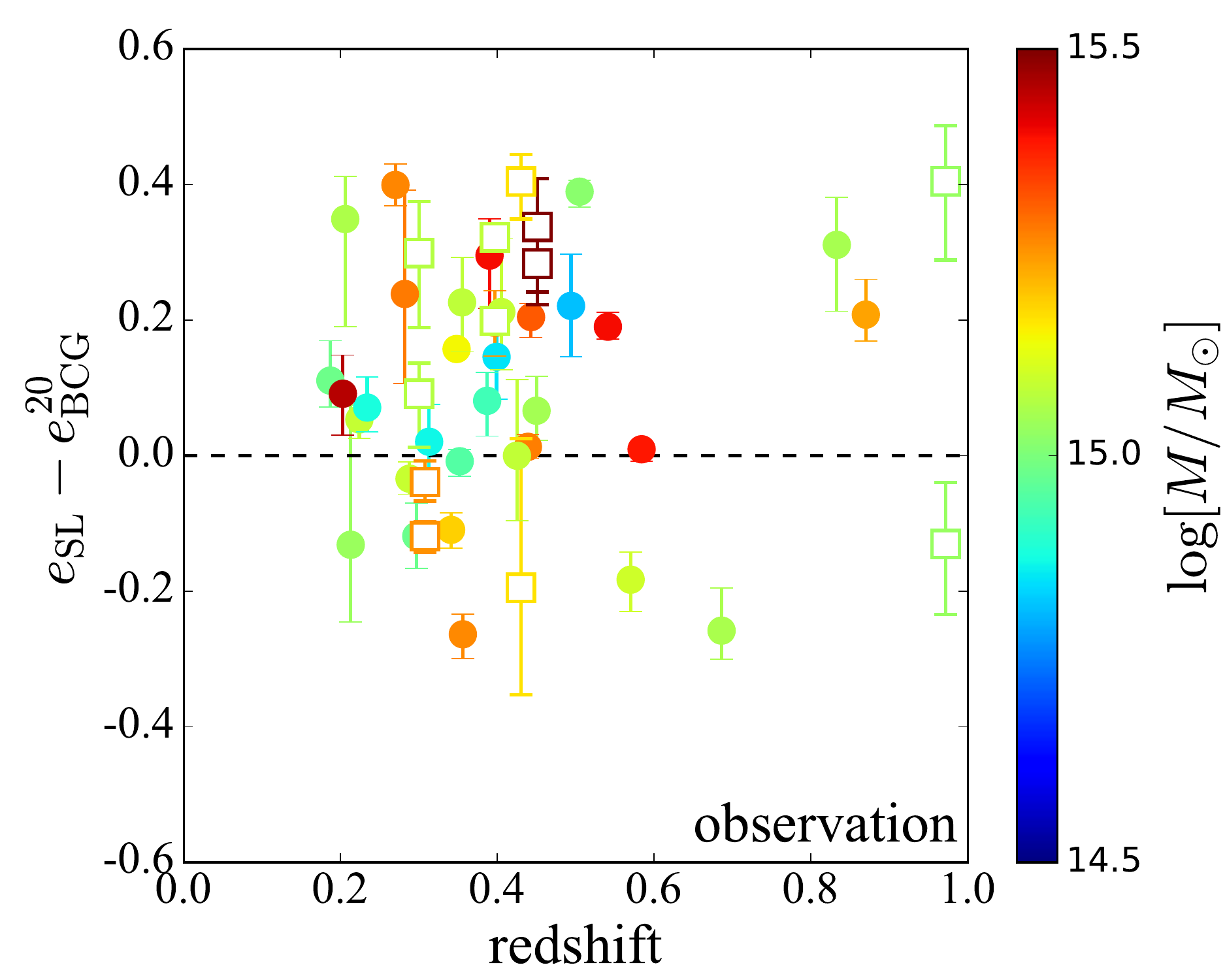}
    \caption{ 
    Differences of ellipticities between DM haloes measured by strong lensing and BCGs fitted at the fiducial scale $R_{ab} = 20$~pkpc as a function of redshift. Symbols are same as in Figure~\ref{fig:ell_sl_bcg_hst}.
    }
    \label{fig:ell_diff_hst}
\end{figure*}

Figures~\ref{fig:ell_sl_z_hst} and \ref{fig:ell_bcg_z_hst} show the redshift dependence of ellipticities of DM haloes and BCGs, respectively, which do not exhibit strong dependence on redshift. Figure~\ref{fig:ell_diff_hst} shows that the difference of ellipticities, $e_{\rm SL} - e_{\rm BCG}^{20}$, which also does not strongly depend on redshift.

\begin{figure*}
\centering
	\includegraphics[width=8.5cm]{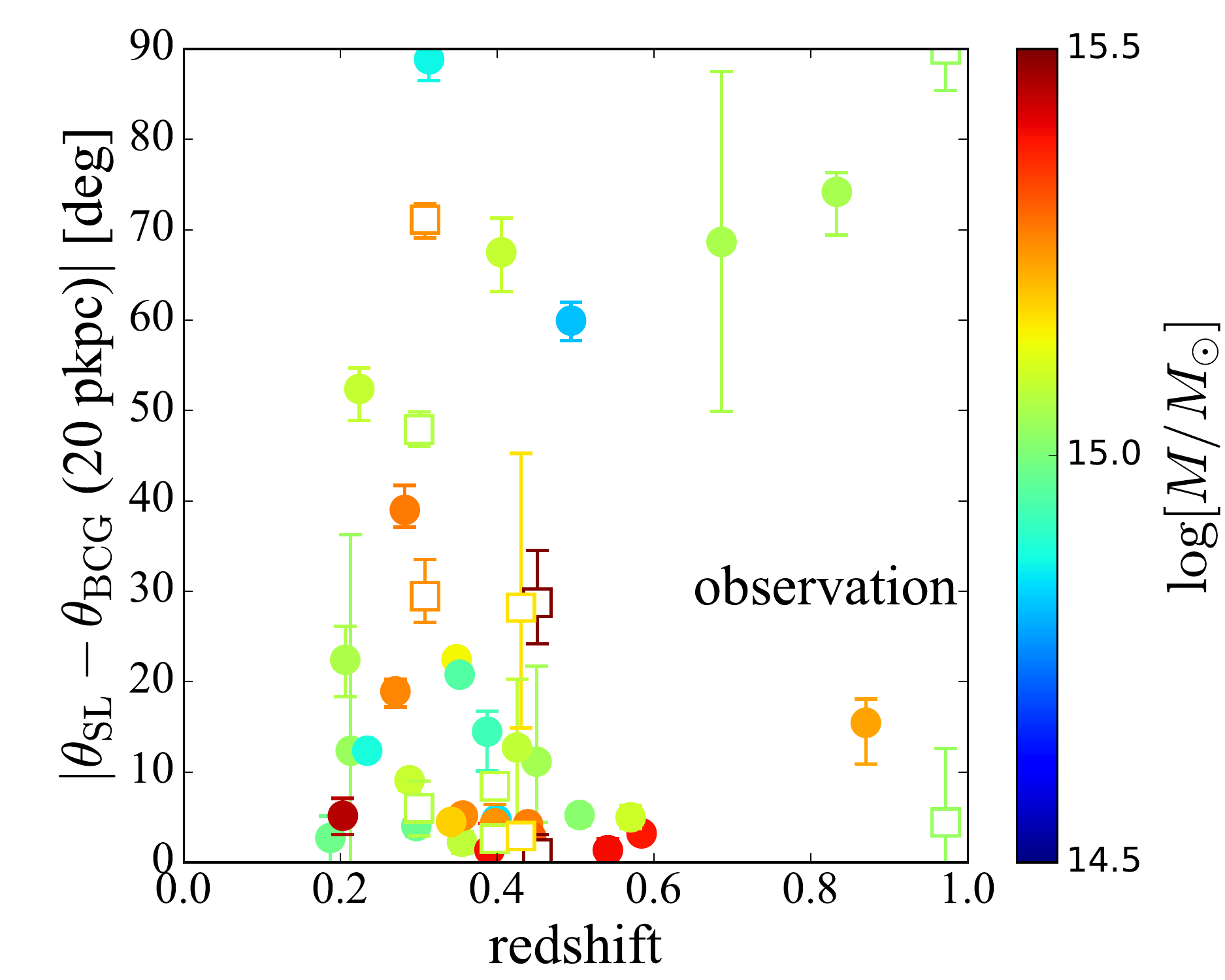}
    \caption{ 
    Alignment angles between DM haloes measured by strong lensing and BCGs fitted at the fiducial scale $R_{ab} = 20$~pkpc as a function of redshift. Symbols are same as in Figure~\ref{fig:ell_sl_bcg_hst}.
    }
    \label{fig:pa_sl_bcg_z_hst}
\end{figure*}

Figure~\ref{fig:pa_sl_bcg_z_hst} shows the redshift dependence of alignment angles between DM haloes and BCGs at $R_{ab} = 20$~pkpc. We find that the alignment angles do not strongly depend on redshift. Since we do not find significant redshift dependence for any observed quantities, we ignore the redshift dependence in the following analysis.

\section{Comparison with Horizon-AGN simulation} \label{sec:sim_obs_ell}
Although we find that ellipticity values of DM haloes are on average larger than those of BCGs in galaxy clusters, they are similar in the Horizon-AGN simulation \citep{okabe18}. One possible explanation is due to the difference of mass scales between observations ($\sim10^{15}M_{\odot}$) and cluster-sized haloes in the Horizon-AGN simulation ($\sim10^{14}M_{\odot}$). In order to check this possibility, we explore the mass dependence of DM haloes and central galaxies (CGs) in the Horizon-AGN simulation. Following \citet{okabe18} and \citet{2019arXiv191104653O}, we refer to CGs as counterparts of BCGs in the simulation. Unfortunately, since there is no DM halo in the Horizon-AGN simulation whose mass is comparable to the observed galaxy clusters, we cannot directly compare the observation with the simulation. Nevertheless, we expect that the analysis of the mass dependence in the Horizon-AGN simulation may provide a clue to the origin of the difference.

In the Horizon-AGN simulation, we identify DM haloes using the ADAPTAHOP halo finder \citep{2004MNRAS.352..376A, 2009A&A...506..647T} and select all DM haloes with masses higher than $10^{12.5}M_{\odot}$, corresponding to massive galaxies. The masses of these haloes are defined by the Friend-of-Friend (FOF) mass which roughly corresponds to the virial mass. Since the redshift dependence of $e$ and $\theta$ both in the observations (see Section~\ref{sec:ell_pa_hst}) and the simulation is weak, we choose a snapshot at redshift $z=0.39$ that is close to the mean value of redshifts of observed clusters, $\langle z \rangle=0.43$. The total number of DM haloes used for the analysis is 1265. In order to make a fair comparison with observations, we create projected particle distributions for each halo. We consider three different projection directions assuming $x$-, $y$-, and $z$-axes as line-of-sight directions and regard these three projections as independent so that we effectively have $N_{\rm cl}\equiv 3795$ DM haloes for our analysis.

We use the mass tensor \citep[see][for details]{okabe18,2019arXiv191104653O} to fit the ellipse to both DM haloes and CGs from simulations. For DM haloes, we use only particles belonging to the most {\it massive} structure in the halo, where substructures are eliminated by ADAPTAHOP finder. This is because in our strong lensing analysis we model the cluster mass distribution by the sum of smooth DM halo components and cluster member galaxies (substructures) and derive DM halo shapes from those of the smooth DM halo components alone (see Section~\ref{sec:method_sl}). In contrast, for the CGs in the simulation, we use all the stellar particles around the central region of the halo in projection because we do not exclude substructures in fitting to the observed BCGs (see Section~\ref{sec:method_bcg}). For the CGs, we extract all the stellar particles in a cube with size of (500~pkpc)$^3$, create project images to use these particles, and perform the ellipse fit. We adopt $R_{ab}=10$, 20, and 30~pkpc for CGs in the same manner as in observations and 100~pkpc for DM haloes that matches the typical Einstein radii of the observed clusters. Even though both typical sizes of CGs and typical Einstein radii are smaller for haloes with smaller masses, we adopt these fixed radii for the ellipse fit independently of the halo mass, because we are mainly interested in the {\it extrapolation} of the simulation result to more massive haloes corresponding to observed clusters.
We note that, while the ellipticity of DM haloes depend on the radius in DM only simulations \cite[e.g.,][]{2006MNRAS.367.1781A}, the radial dependence is found to be much weaker in hydrodynamical simulations \cite[e.g.,][]{okabe18}. Thus in this paper we  ignore the radial dependence of the shape of DM haloes for simplicity.
Since the spatial resolution of the Horizon-AGN simulation of $\sim1$~kpc is sufficiently small compared with the ellipse scales, we can safely ignore the effect of the spatial resolution in our analysis.

\begin{figure*}
\centering
	\includegraphics[width=8.0cm]{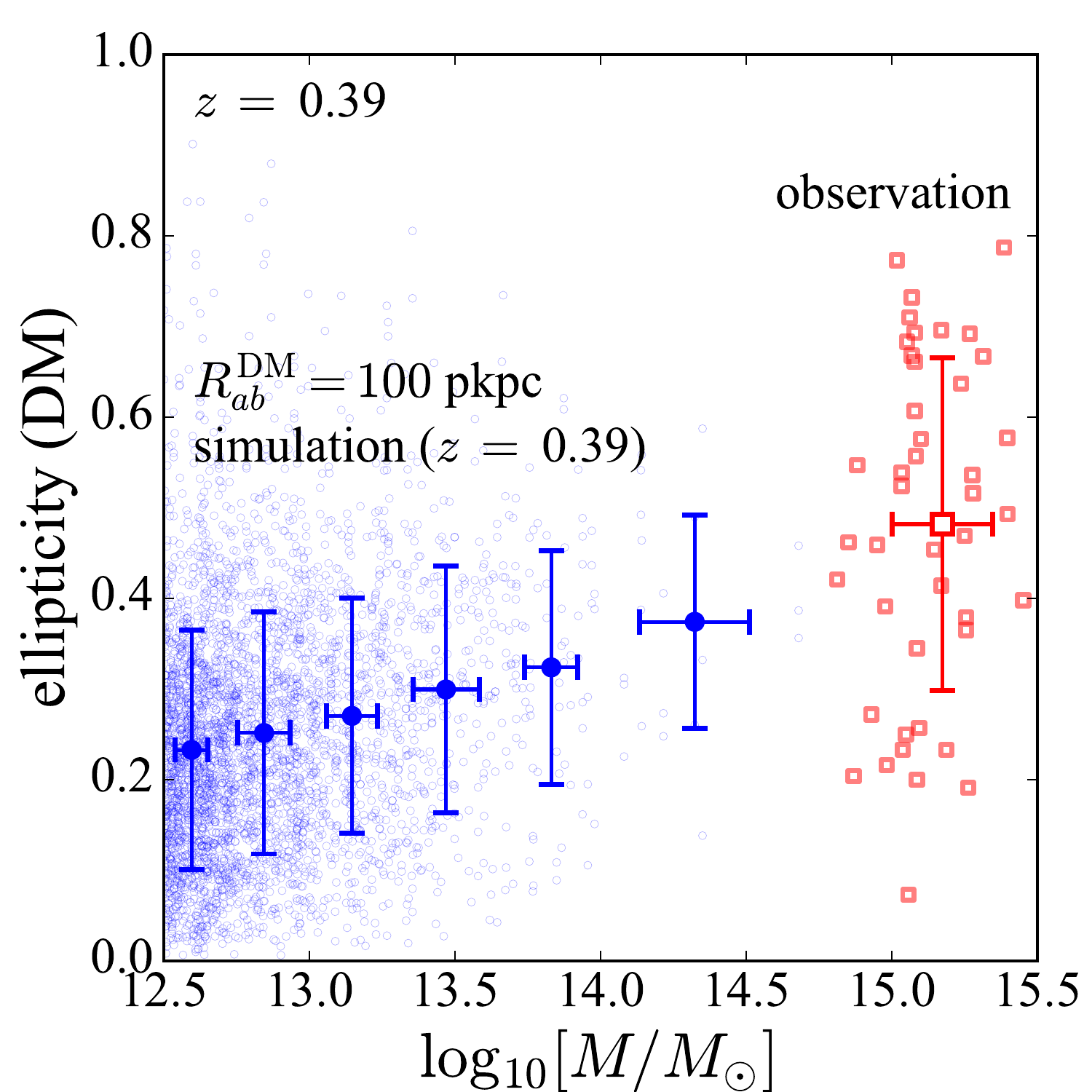}
    \caption{ 
    Large filled circles and open square show mean values of ellipticities of DM haloes derived from the Horizon-AGN simulation and strong lensing observations, respectively. We adopt redshift $z=0.39$ and fitted ellipse scale $R_{ab}=100$~pkpc in the Horizon-AGN simulation. Error-bars of $x$- and $y$-axis directions denote the standard deviation of DM halo mass and ellipticities, respectively. Small circles and squares show ellipticities of individual DM haloes in the Horizon-AGN simulation and observations, respectively.
    }
    \label{fig:esl_mass_mean}
\end{figure*}

After we fit the ellipses for each halo in the simulation by the above procedure, we divide the haloes in 6 bins according to their DM halo masses, $M<5\times10^{12} M_{\odot}$, $5\times10^{12}M_{\odot}\leq M<10^{13} M_{\odot}$, $10^{13}M_{\odot}\leq M<2\times10^{13} M_{\odot}$, $2\times10^{13}M_{\odot}\leq M<5\times10^{13} M_{\odot}$, $5\times10^{13}M_{\odot}\leq M<10^{14} M_{\odot}$, and $10^{14}M_{\odot} \leq M$. We then compute mean values of ellipticities and alignment angles between DM haloes and CGs for each mass bin. Figure~\ref{fig:esl_mass_mean} shows mean values of ellipticities of DM haloes as a function of DM halo mass. There is a clear trend that shapes of DM haloes with higher mass are on average more elliptical than those with smaller masses. This result is qualitatively consistent with the result in e.g., \citet{2014MNRAS.443.3208D}, who analysed three different cosmological simulations and investigated mass dependence of halo shapes. While their shape measurement is based on the three dimensional triaxial fitting and thus cannot be directly compared with our results, these DM only simulations also indicate that haloes with higher masses are more elongated (see their Figure~4). This is presumably because more massive DM haloes are dynamically young and still experiencing major mergers or smooth mass accretions along filaments \citep[see also][]{2019arXiv191104653O}, whereas less massive ones form at earlier epochs and thus they have enough time to reach relaxation. 

Mean value of ellipticities of observed clusters is also plotted in Figure~\ref{fig:esl_mass_mean}. As expected, the observed value ($\langle e_{\rm DM }\rangle=0.482\pm 0.028$) is higher than those of DM haloes in the simulation, suggesting that the mass dependence of ellipticities might explain why $e_{\rm DM } > e_{\rm BCG}$ for the observed clusters. Table~\ref{tab:mean_values_of_ell_sl} shows mean values of ellipticities of DM haloes. We also compute mean values of ellipticities for single- and double-peak clusters and find that double-peak clusters are more elongated than single-peak clusters. This is naturally understood because double-peak clusters are expected to be dynamically younger on average than single-peak clusters.
\begin{table}
	\centering
	\caption{Mean values and their errors of ellipticities of DM haloes. The errors are defined as standard deviation divided by the square root of number of DM haloes in each bin. }
	\label{tab:mean_values_of_ell_sl}
	\begin{tabular}{ l l l l } 
		\hline
		\hline 
		  & & $\log(\langle M_{\rm DM}\rangle)$ & $\langle e_{\rm DM}\rangle$ \\
		  & & [$\log(M_{\odot})$] & \\
                  \hline 
		  observation & all & $15.17$ & $0.482\pm0.028$ \\
		       & single peak & $15.15$ & $0.451\pm0.033$ \\
		      & double peak & $15.23$ & $0.567\pm0.040$ \\ 
		\hline
		      & HFF & $15.22$ & $0.507\pm0.050$ \\ 
		      & CLASH & $15.14$ & $0.418\pm0.046$ \\
		      & RELICS & $15.19$ & $0.535\pm0.036$ \\ 
		\hline
		simulation && $12.6$ & $0.233\pm0.003$ \\
		                && $12.8$ & $0.252\pm0.004$ \\
		                && $13.1$ & $0.270\pm0.005$ \\
				  && $13.5$ & $0.300\pm0.008$ \\
	   			  && $13.8$ & $0.324\pm0.015$ \\
				  && $14.3$ & $0.374\pm0.026$ \\
		\hline 
	\end{tabular}
\end{table}

\begin{figure*}
\centering
	\includegraphics[width=8.0cm]{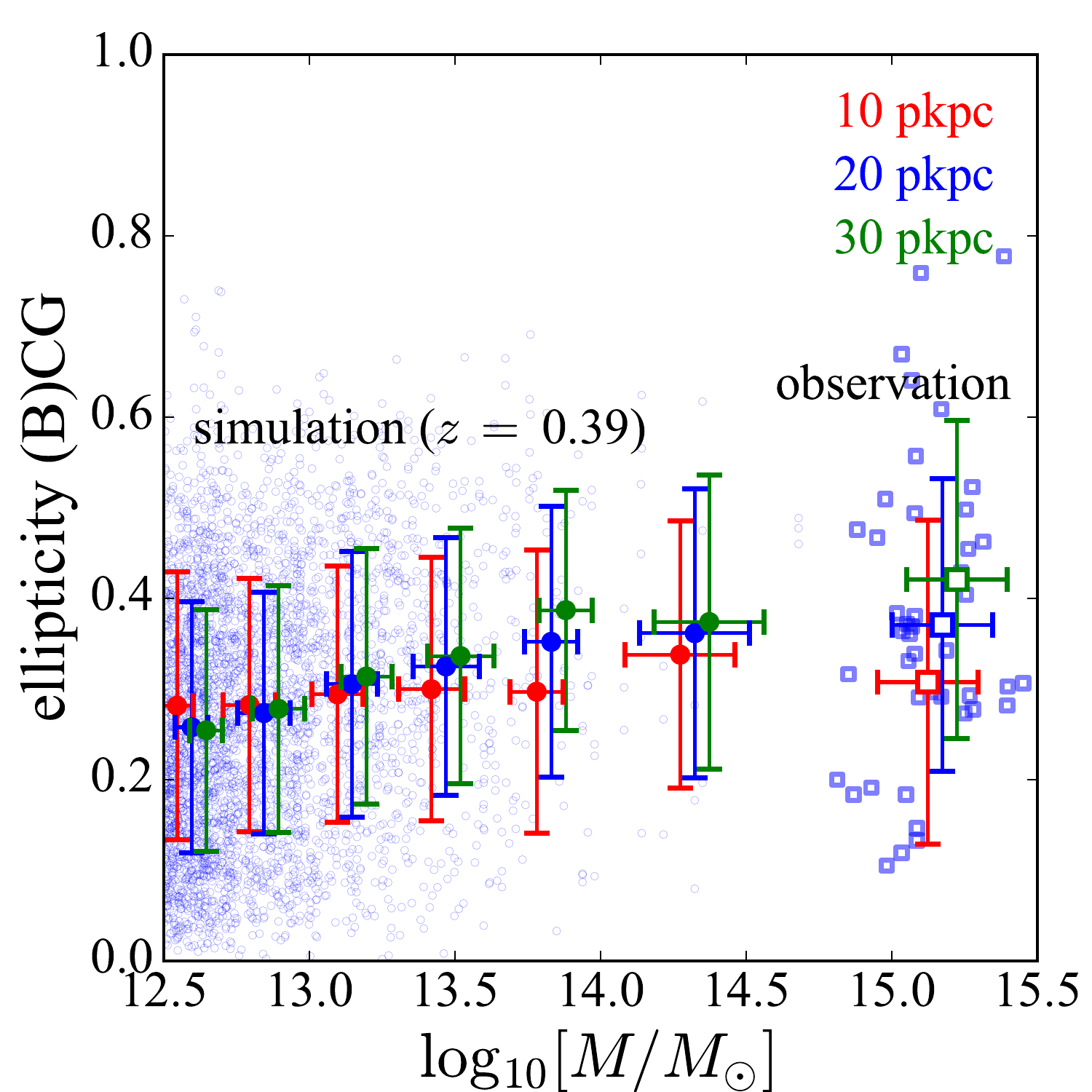}
    \caption{ 
    Large filled circles and open squares show mean values of ellipticities of CGs derived from the Horizon-AGN simulation and BCGs in {\it HST} observations, respectively. We show results for different ellipse scales, $R_{ab}= 10$ (red), 20 (blue), and 30 (green) pkpc, respectively, for both observation and simulation. Just for the clarity, red and green symbols are shifted by $-0.05$ and $+0.05$ in the horizontal direction, respectively. We adopt redshift $z=0.39$ for the analysis of the Horizon-AGN simulation. Error-bars of $x$- and $y$-axis directions denote the standard deviation of DM halo mass and ellipticities, respectively. Small circles and squares are ellipticities of individual CGs of the Horizon-AGN simulation and BCGs of {\it HST} observations at $R_{ab}=20$~pkpc, respectively. }
    \label{fig:ebcg_mass_mean}
\end{figure*}

Figure~\ref{fig:ebcg_mass_mean} plots mean values of ellipticities of BCGs in observations and CGs in the simulation as a function of DM halo mass. For the scales of $R_{ab} = 20$ and 30~pkpc, we find the trend similar to DM haloes, whereas for 10~pkpc, mean values of ellipticities are almost constant against the halo mass in the simulation. For lower halo mass, the CG shapes are rounder at larger scales, whereas for higher mass, these are more elongated at larger scales. One possible reason of this result is that inner regions formed at the earlier epoch and have enough time to relax, and thus they are not affected by accretion or formation history and are independent of the host halo mass. Another possibility is the effect of satellite galaxies that tend to exist at larger scales and make ellipse more elongated.

Since we adopt the same tensor method for ellipse fit of observed BCGs and simulated CGs, they can be more directly compared unlike DM haloes. Figure~\ref{fig:ebcg_mass_mean} suggests that observed values can well be explained by the {\it extrapolation} of the simulation.  Table~\ref{tab:mean_values_of_ell_sl} shows mean values of ellipticities of observed BCGs and CGs in the simulation. While double-peak clusters are more elongated than single-peak clusters in the outer region, 30~pkpc, their values are similar at 10~pkpc. This is presumably because stellar components in the inner region are tightly bound with each other, and thus their distributions are not affected by major mergers or mass accretions. Another possible explanation is that satellite galaxies are likely to be more abundant in double-peak clusters as they are dynamically younger and hence shape measurements of BCGs in the outer region are more severely affected by satellite galaxies.
\begin{table}
	\centering
	\caption{ Mean values and their errors of ellipticities of (B)CGs. The errors are defined as standard deviation divided by the square root of number of (B)CGs in each bin. }
	\label{tab:mean_values_of_ell_bcg}
	\begin{tabular}{ l l l l l } 
		\hline
		\hline 
		  & & $\log(\langle M_{\rm DM}\rangle)$ &
                $R_{ab}$ & $\langle e_{\rm (B)CG}\rangle$ \\
                  & & [$\log(M_{\odot})$]& [pkpc] & \\
		  \hline 
		  observation & all & $15.17$ &10 & $0.308\pm0.027$\\
		                                                    &&& 20 & $0.370\pm0.024$  \\
		                          			           &&& 30 & $0.421\pm0.026$ \\ 
		         &single peak & $15.15$ & 10 & $0.305\pm0.030$   \\  
		              					    &&& 20 & $0.355\pm0.027$ \\
								           &&& 30 & $0.399\pm0.029$ \\
			   &double peak & $15.23$ & 10 & $0.314\pm0.062$ \\	
			                                             &&& 20 & $0.412\pm0.052$ \\
			                                             &&& 30 & $0.480\pm0.057$ \\  
		\hline 
		  & HFF & $15.22$ &10 & $0.290\pm0.073$\\
		                        &&& 20 & $0.389\pm0.034$  \\
		                        &&& 30 & $0.366\pm0.063$ \\ 
		  & CLASH & $15.14$ &10 & $0.240\pm0.026$\\
		                            &&& 20 & $0.345\pm0.043$  \\
		                            &&& 30 & $0.361\pm0.037$ \\   
		  & RELICS & $15.19$ &10 & $0.381\pm0.044$\\
		                            &&& 20 & $0.389\pm0.033$  \\
		                            &&& 30 & $0.494\pm0.038$ \\             
		\hline 
		simulation && $12.6$ & 10 & $0.282\pm0.004$ \\
		                                           &&& 20 & $0.258\pm0.003$ \\
		                                           &&& 30 & $0.254\pm0.003$ \\
				  && $12.8$ & 10 & $0.282\pm0.004$ \\
				                             &&& 20 & $0.274\pm0.004$ \\
				                             &&& 30 & $0.278\pm0.004$ \\
				  && $13.1$ & 10 & $0.294\pm0.006$ \\
				                             &&& 20 & $0.305\pm0.006$ \\
				                             &&& 30 & $0.314\pm0.006$ \\
				  && $13.5$ & 10 & $0.300\pm0.008$ \\
				                             &&& 20 & $0.325\pm0.008$ \\
				                             &&& 30 & $0.336\pm0.008$ \\
				  && $13.8$ & 10 & $0.297\pm0.018$ \\
				                             &&& 20 & $0.352\pm0.017$ \\
				                             &&& 30 & $0.387\pm0.015$ \\
				  && $14.3$ & 10 & $0.338\pm0.032$ \\
				                             &&& 20 & $0.362\pm0.035$ \\
				                             &&& 30 & $0.374\pm0.035$ \\ 
		\hline 
	\end{tabular}
\end{table}

\begin{figure*}
\centering
	\includegraphics[width=8.0cm]{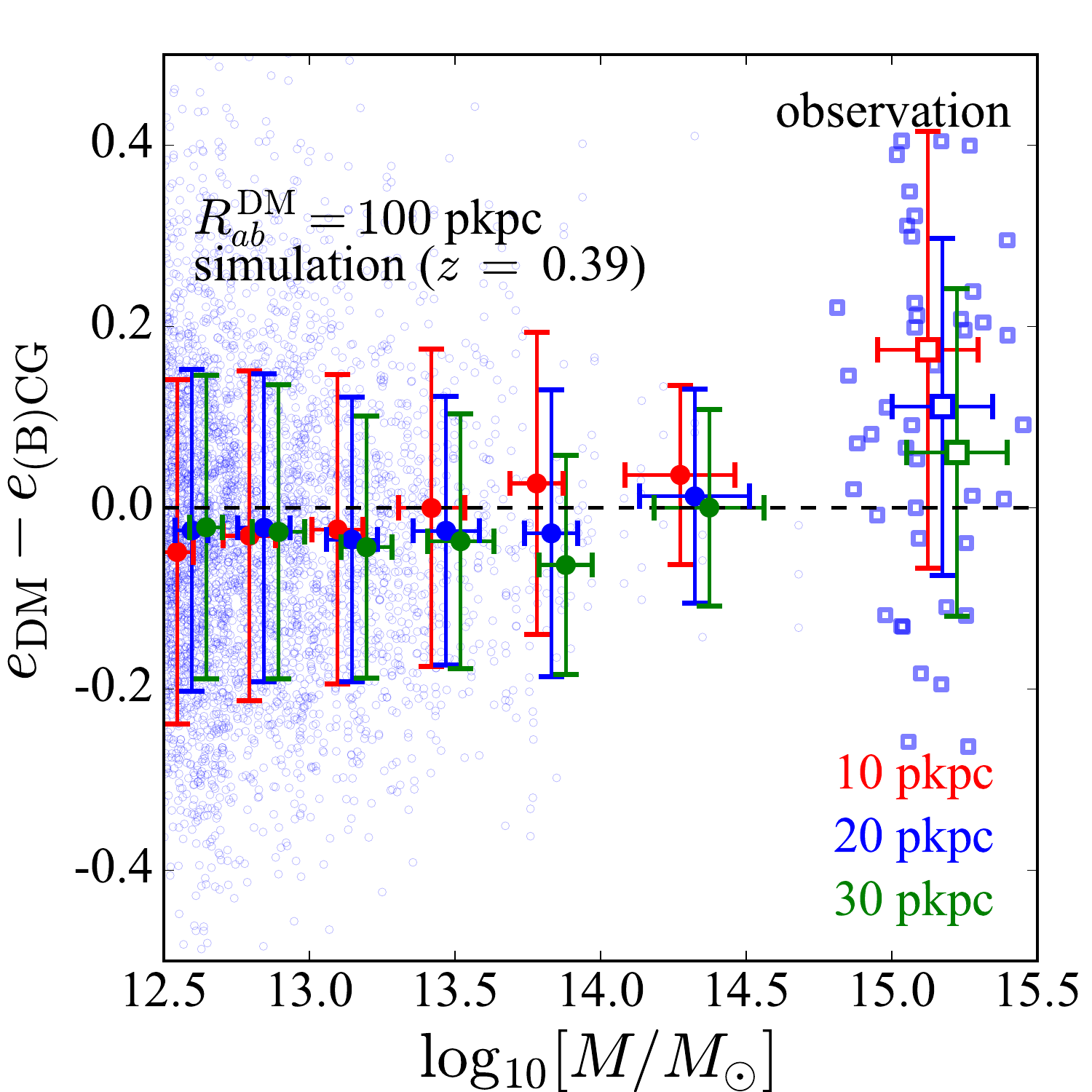}
    \caption{Mean values of difference between ellipticities of DM haloes and those of (B)CGs. Symbols are same as in Figure~\ref{fig:ebcg_mass_mean}.
    }
    \label{fig:esl_ebcg_mass_mean}
\end{figure*}

Figure~\ref{fig:esl_ebcg_mass_mean} plots mean values of differences between ellipticities of DM haloes and those of observed BCGs and simulated CGs. While the mean values are close to zero in the simulation, those of observed values are significantly higher than zero, $0.1$--$0.2$. As we discussed in Section~\ref{sec:ell_pa_hst}, this difference might be due to the difference of mass scales between observations and the simulation. Figure~\ref{fig:esl_ebcg_mass_mean} suggests that there is no strong trend of the mean values against the halo mass in the simulation. There is, however, a weak trend of increasing $\langle e_{\rm SL}-e_{\rm BCG} \rangle$ particularly for $R_{ab}=10$~pkpc, which might explain observed values by extrapolating the mass dependence.
Since the mean ellipticies of $e^{10}_{\rm BCG}$ are almost constant with mass and mean ellipticities of $e_{\rm SL}$ are not, the trend mainly comes from the mass dependence of $e_{\rm SL}$.

\begin{figure*}
\centering
	\includegraphics[width=8.0cm]{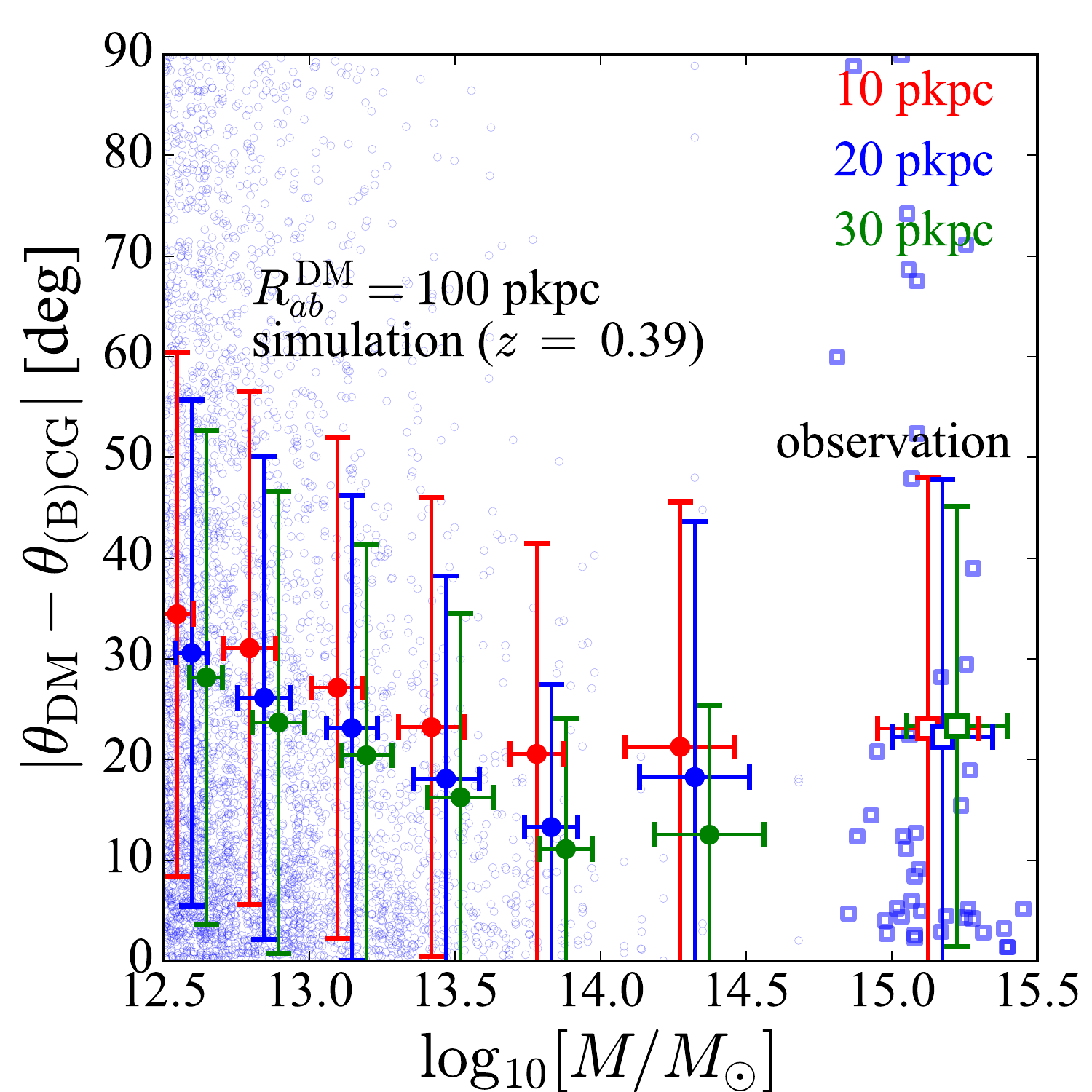}
    \caption{
    Mean values of alignment angles between major axes of DM haloes and those of (B)CGs. Symbols are the same in Figure~\ref{fig:ebcg_mass_mean}.
    }
    \label{fig:theta_mean_mass}
\end{figure*}

In addition to the mass dependence of ellipticities, we investigate that of alignment angles between DM haloes and CGs. Figure~\ref{fig:theta_mean_mass} plots mean values of the alignment angles. In the low mass region $\log[M/M_{\odot}]<14.0$, there is a clear trend that the alignment becomes tighter with increasing masses. However, in the high mass region $\log[M/M_{\odot}]>14.0$, the alignment appears to be constant independent of the mass increase. Table~\ref{tab:mean_values_of_theta_align} summarizes mean values of the alignment angles.
\begin{figure*}
\centering
	\includegraphics[width=8.0cm]{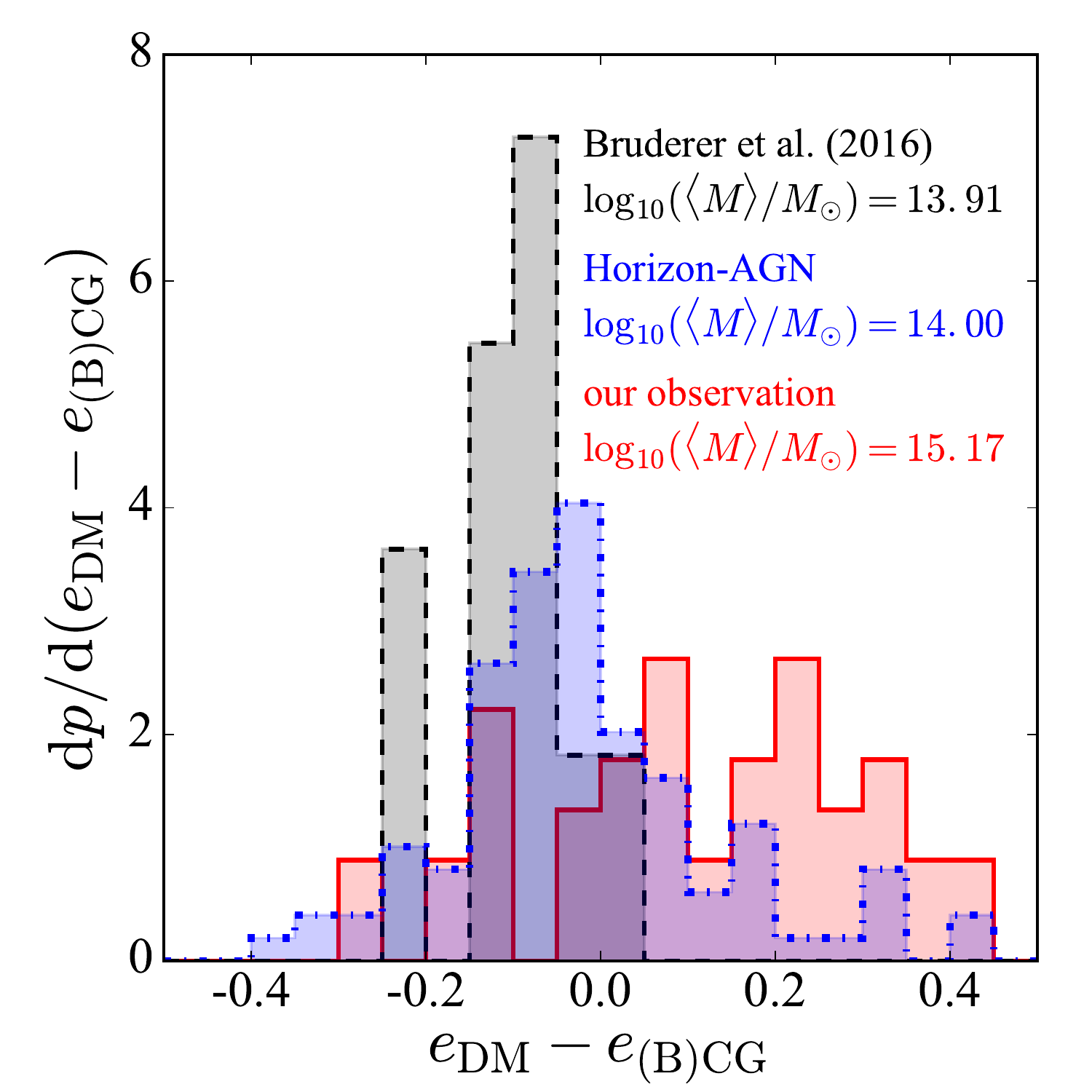}
    \caption{
    The probability distributions of the ellipticity difference, $e_{\rm DM}-e_{\rm (B)CG}$, observed by \citet{2016MNRAS.456..870B} (black dashed), our observation (red solid), and in the Horizon-AGN simulation (blue dot-dashed).
    We use only haloes with their masses larger than $5\times10^{13}M_{\odot}$ in the Horizon-AGN simulation.
    }
    \label{fig:dp_de_mean_diff}
\end{figure*}

\begin{table}
	\centering
	\caption{Mean values and their errors of alignment angles between DM haloes and the CGs. The errors are defined as standard deviation divided by the square root of number of DM haloes in each bin. }
	\label{tab:mean_values_of_theta_align}
	\begin{tabular}{ l l l l l } 
		\hline
		\hline 
		  & & $\log(\langle M_{\rm DM}\rangle)$ & $R_{ab}$ & $\langle |\theta_{\rm DM}-\theta_{\rm (B)CG}|\rangle$\\
                  & & [$\log(M_{\odot})$] & [pkpc] &  [deg] \\
		\hline
		  observation & all & $15.17$ & 10 & $23.1\pm3.8$  \\
		                                                   &&& 20 & $22.2\pm3.9$ \\
		                                                   &&& 30 & $23.3\pm3.3$ \\ 
		       & single peak & $15.15$ & 10 & $22.3\pm4.6$ \\
		                                                   &&& 20 & $20.6\pm4.3$ \\
		                                                   &&& 30 & $21.8\pm3.7$ \\
		      & double peak & $15.23$ &10 & $25.2\pm6.3$ \\ 
		      					                 &&& 20 & $26.7\pm8.5$ \\
							                 &&& 30 & $27.4\pm7.0$ \\ 
		 \hline 
		   & HFF & $15.22$ & 10 & $37.6\pm8.7$  \\
		                          &&& 20 & $22.6\pm9.8$ \\
		                          &&& 30 & $21.3\pm10.0$ \\ 
		       & CLASH & $15.14$ & 10 & $16.7\pm4.8$ \\
		                                  &&& 20 & $22.3\pm5.8$ \\
		                                  &&& 30 & $24.0\pm5.3$ \\
		      & RELICS & $15.19$ &10 & $24.8\pm6.0$ \\ 
		      				     &&& 20 & $22.0\pm5.9$ \\
						     &&& 30 & $23.2\pm4.3$ \\ 
		\hline 
		simulation && $12.6$ & 10 & $34.4\pm0.7$ \\
								&&& 20 & $30.6\pm0.6$ \\
								&&& 30 & $28.2\pm0.6$ \\
				&& $12.8$ & 10 & $31.0\pm0.7$ \\
								&&& 20 & $26.1\pm0.7$ \\
								&&& 30 & $23.7\pm0.7$ \\
				&& $13.1$ & 10 & $27.1\pm1.0$ \\
								&&& 20 & $23.1\pm1.0$ \\
								&&& 30 & $20.4\pm0.9$ \\
				&& $13.5$ & 10 & $23.2\pm1.3$ \\
								&&& 20 & $18.1\pm1.1$ \\
								&&& 30 & $16.3\pm1.0$ \\
				&& $13.8$ & 10 & $20.6\pm2.4$ \\
								&&& 20 & $13.3\pm1.6$ \\
								&&& 30 & $11.1\pm1.5$ \\
				&& $14.3$ & 10 & $21.3\pm5.3$ \\
								&&& 20 & $18.3\pm5.5$ \\
								&&& 30 & $12.5\pm2.8$ \\
		\hline 
	\end{tabular}
\end{table}

\section{Discussion} \label{sec:discussions_comparison_obs}
We find that the mean difference of ellipticities between DM haloes and BCGs in observations is $\langle e_{\rm SL} - e_{\rm BCG}\rangle = 0.11\pm0.03$. 
This seems inconsistent with the Horizon-AGN simulation results for which the mean value of the ellipticity difference is $\langle e_{\rm SL} - e_{\rm CG}\rangle = -0.020\pm0.015$ with DM halo mass of $M_{\rm DM}>5\times10^{13}M_{\odot}$.
We consider several possibilities to explain the difference, which are discussed below.

First, as already mentioned, a possible explanation comes from the difference of mass scales. Figure~\ref{fig:esl_mass_mean} indicates that ellipticities of DM haloes show a clear trend with mass and the observed value might be explained by the {\it extrapolation} of values in the simulation. Figure~\ref{fig:ebcg_mass_mean} shows that the observed ellipticity values of the BCGs can be explained by the extrapolation of the simulation, and thus the observed difference could also be explained by the mass dependence. Figure~\ref{fig:esl_ebcg_mass_mean} indicates that the difference of ellipticities $e_{\rm DM}-e_{\rm CG}$ in the simulation shows a weak trend especially at the inner region such that the extrapolation of the trend may explain the observation. The possibility of this mass dependence may also be tested by other observations at smaller masses.
Figure~\ref{fig:dp_de_mean_diff} compares the probability distributions of the ellipticity difference for our observation and the Horizon-AGN simulation with that of previous observational work by \citet{2016MNRAS.456..870B}, in which they measure projected shapes of 11 DM haloes by strong lensing and compare them with those of light profiles of the central galaxies. Since their definition of the ellipticity $(a^2-b^2)/(a^2+b^2)$ is different from ours, $1-b/a$ with $a$ and $b$ being lengths of semi-major and -minor axes, respectively, we convert their values to our definition.
Their results show the opposite trend $e_{\rm SL}<e_{\rm CG}$, implying that the mass dependence is strong \citep[see also][for a similar result]{2016MNRAS.458....2R}, although a caveat is that their strong lensing measurements probe radii smaller than 100~pkpc that we adopted in the simulation. Figure~\ref{fig:dp_de_mean_diff} also indicates that the probability distribution of the ellipticity difference in \citet{2016MNRAS.456..870B} differs from that in the Horizon-AGN simulation with similar halo masses.
More strong lens samples at different mass scales as well as simulations in larger box sizes are required to test this scenario further.

Second, another possibility is that the strong lensing method we use to measure ellipticities of DM haloes is biased such that it derives higher ellipticity values than those of real DM mass distributions. To check this possibility, Figure~\ref{fig:ell_umetsu} compares our measurement values by strong lensing with those by weak lensing analysis \citep{2018ApJ...860..104U} for 15 galaxy clusters whose ellipticities are evaluated by both strong and weak lensing, although a caveat is that strong and weak lensing probe different radii.  The mean value estimated by strong lensing, $\langle e_{\rm SL}\rangle = 0.405\pm0.053$, is higher than those by weak lensing, $\langle e_{\rm WL} \rangle =0.344\pm0.04$, although they are still consistent with each other within the errors. On the other hand, the comparison of position angles shown in Figure~\ref{fig:theta_umetsu} indicates that both position angles are well aligned with each other despite the large errors for weak lensing measurements. Although we cannot draw any robust conclusion because weak and strong lensing measure ellipticities at different scales \citep[see also e.g.,][for misalignment between inner and outer DM haloes]{2010MNRAS.405.2215O,2019arXiv191106333H}, this result implies that the strong lensing method might slightly over-estimates ellipticities.

On the other hand, \citet{2017MNRAS.472.3177M} compares real DM mass distributions with those inferred from various strong lensing methods by using simulated cluster images with mock multiple images which mimic the Hubble Frontier Field survey. This mock challenge demonstrated that if there are a sufficient number of multiple images (say $N_{\rm img}>100$), strong lensing method accurately reproduces input DM mass distributions. In fact, our lensing method is one of the best methods to reproduce shapes of simulated haloes (see ``GLAFIC'' panel of their Figure~7). However, there are not many multiple images for some of the observed clusters (see Appendix~\ref{sec:strong_lensing_method}), for which derived ellipticities might be biased. 

\begin{figure*}
\centering
	\includegraphics[width=0.9\columnwidth]{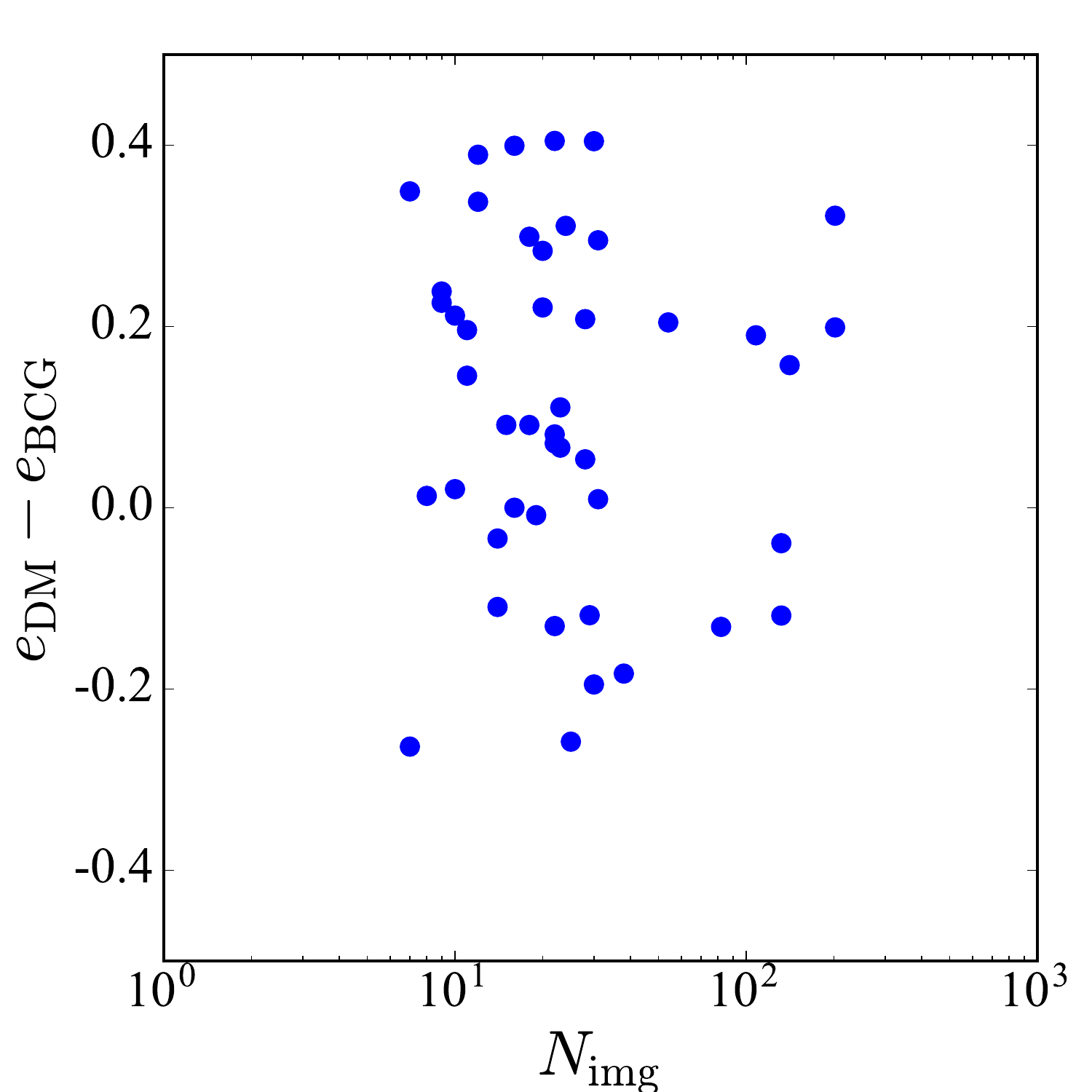}
    \caption{
    The comparison between the numbers of multiple images used for strong lens mass modeling (see Table~\ref{tab:slens}) and ellipticity differences, $e_{\rm DM}-e_{\rm BCG}$.
    }
    \label{fig:ediff_nimg}
\end{figure*}

\begin{figure*}
\centering
	\includegraphics[width=0.9\columnwidth]{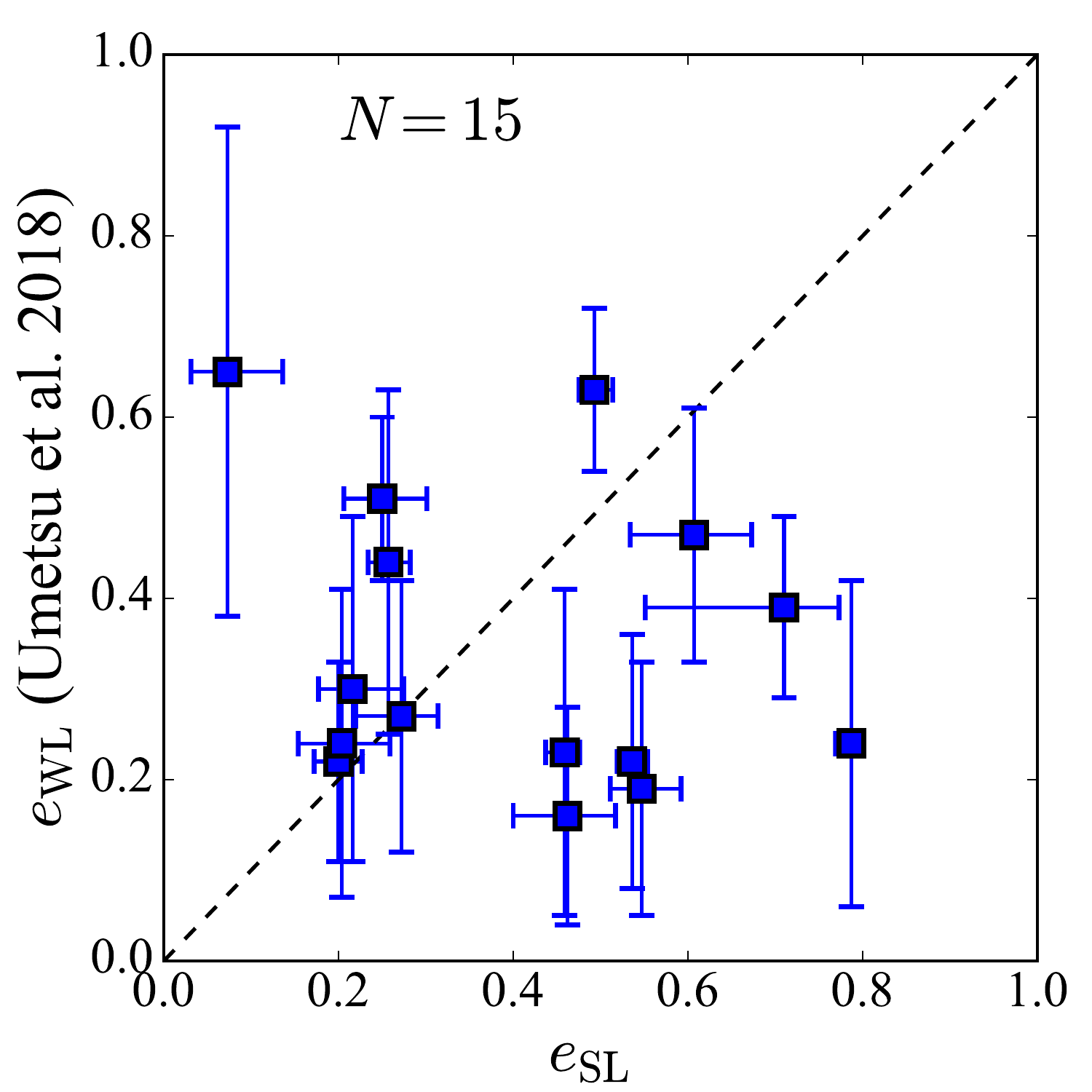}
    \caption{Correlation between values of ellipticities measured by strong lensing ($x$-axis) in this paper and those by weak lensing ($y$-axis) from \citet{2018ApJ...860..104U} for 15 galaxy clusters whose ellipticities are measured by both methods.}
    \label{fig:ell_umetsu}
\end{figure*}

\begin{figure*}
\centering
	\includegraphics[width=0.9\columnwidth]{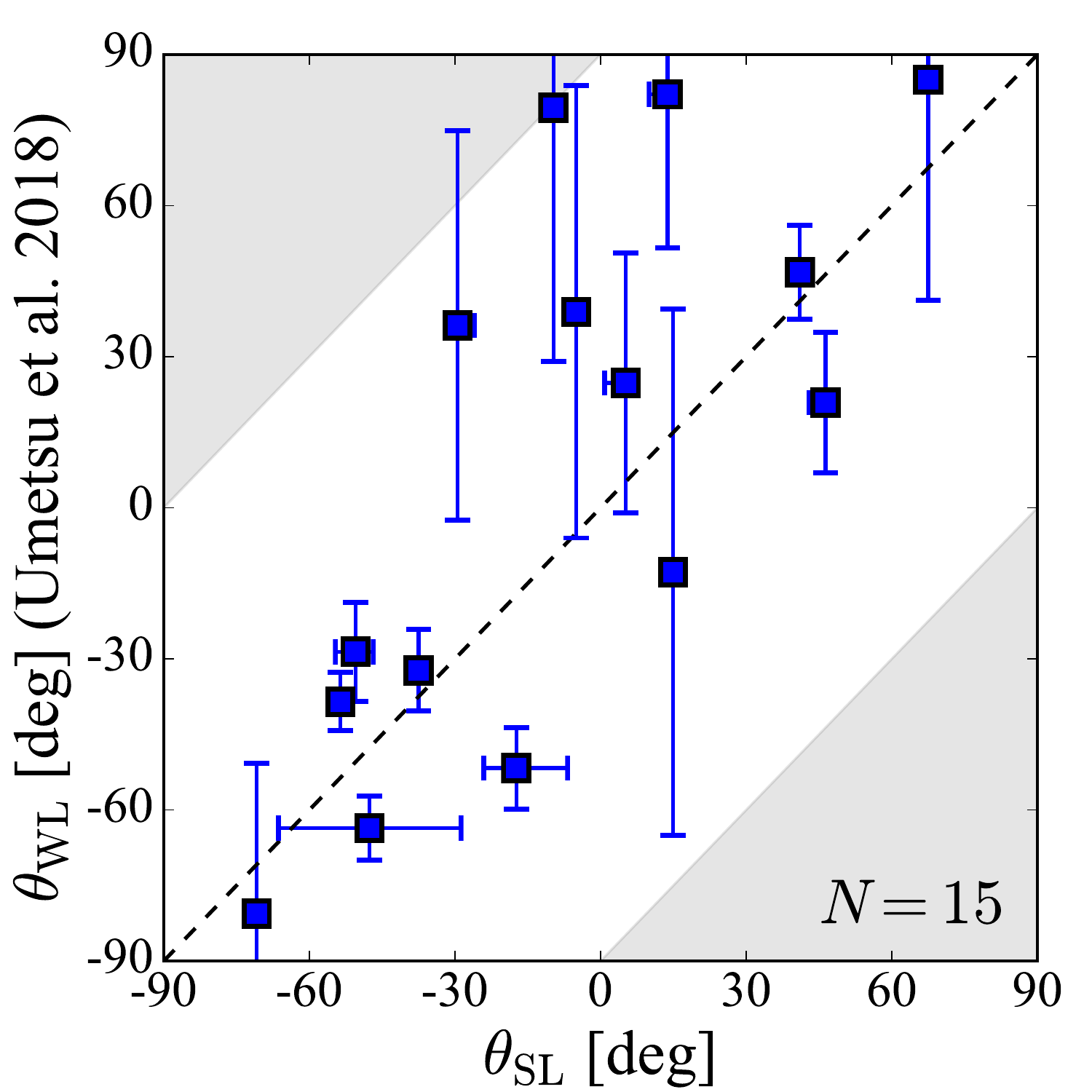}
    \caption{Similar to Figure~\ref{fig:ell_umetsu}, but for position angles. See the caption of Figure~\ref{fig:theta_sl_bcg_hst} for the explanation of the shaded regions.}
    \label{fig:theta_umetsu}
\end{figure*}

To check this possibility, in Figure~\ref{fig:ediff_nimg} we compare ellipticity differences $e_{\rm DM}-e_{\rm BCG}$ with the number of multiple images used for strong lens mass modeling. While the lack of any systematic correlation in Figure~\ref{fig:ediff_nimg} suggests that such bias in measured ellipticity values from strong lensing may not be significant, future studies to validate strong lensing methods to measure ellipticities are warranted \citep[see][for another validation test]{2019arXiv191106333H}.
Another possible systematic effect comes from the number of dark matter haloes that is allowed to vary in strong lens mass modeling. In the GLAFIC mass modeling, the number of haloes corresponds to the minimum number of haloes that leads to a reasonable fit to multiple image positions \citep[see][]{2016ApJ...819..114K}, but such ellipticity measurements may be biased if the number of halos is not corresponding to the underlying mass distribution  \cite[see e.g.,][]{2017MNRAS.469.3946L}. This issues should also be explored carefully in validating strong lensing methods to measure ellipticities.

Finally, there are also some other possible explanations. One is the difference of ellipticities between DM haloes and BCGs might be explained by selection effects.
For instance, all the clusters in HFF and a small fraction of clusters in CLASH and RELICS are selected based on their strong lensing features. Since projected mass distributions of clusters having large Einstein radii are rounder because they are preferentially elongated along the line-of-sight direction \citep{2009MNRAS.392..930O}, this selection effect might also affect the statistics of the ellipticity difference studied in this paper. In addition, a large fraction of the CLASH clusters are selected such that their shapes are round in X-ray images, which also represents a biased cluster population.
Although Tables~\ref{tab:mean_values_of_ell_sl} and \ref{tab:mean_values_of_ell_bcg} show that mean values of ellipticities of CLASH clusters are indeed smaller than in the other surveys, we find that differences of ellipticities between DM haloes and BCGs for these three surveys are consistent within error-bars, $0.118\pm0.061$ (HFF), $0.072\pm0.037$ (CLASH), and $0.146\pm0.046$ (RELICS). Therefore selection effects do not provide convincing explanation for our finding, although it is important to check ellipticities of DM haloes and BCGs for several different cluster samples to strengthen our conclusion on the measurements.

It is also possible that the Horizon-AGN simulation produces DM haloes or CGs with their shapes that are different from their true shapes in observations due to an imperfect modeling of baryon physics. Although the Horizon-AGN simulation is successful in explaining various observations \citep{2015MNRAS.454.2736C,2016MNRAS.463.3948D,2016MNRAS.460.2979V,2016MNRAS.461.2702C,2017MNRAS.467.4739K,2017MNRAS.472.2153P,suto17,2018MNRAS.480.3962C,2018MNRAS.474.3140M,2018ApJ...856..114C,okabe18,2019MNRAS.483.4615P,2019MNRAS.489.1206H}, the implemented baryon physics is never perfect and the change of details of baryon physics may change quantitative results on halo shapes in simulations \citep[see e.g.,][and references therein]{suto17}. Turning the problem around, we may be able to test the baryon physics such as AGN feedback by observations of ellipticities \citep[see also][]{2012ApJ...755..116L}.

The remaining possibility is that the $\Lambda$CDM model is not correct. Although the standard $\Lambda$CDM model has passed many observational tests there remains several challenges at small-scales that need to be resolved \citep[e.g.,][]{2017ARA&A..55..343B}. For example, the self-interacting dark matter (SIDM) model is proposed as one of the possibilities to solve small-scale problems \citep[e.g.,][]{2000PhRvL..84.3760S,2018PhR...730....1T}, although \citet{2019MNRAS.488.3646R} investigate halo shapes by using cosmological simulations including both baryon physics and SIDM to show that the difference of ellipticities between collisionless and SIDM haloes become larger in the inner region such that SIDM haloes are on average rounder. Therefore it appears that SIDM cannot reconcile the difference between observations and Horizon-AGN simulation, but there may be other DM scenarios that better explain the observations.

While we cannot draw any robust conclusion on which scenario is correct, our observations can be regarded as new constraints on structure formation scenarios including dark matter models, theories of modified gravity, and cluster astrophysics.

\section{Summary} \label{sec:summary_comparison_obs}
In this paper, we have presented new measurements of ellipticities and position angles of galaxy clusters selected from three deep imaging surveys by {\it HST}, HFF, CLASH, and RELICS. The measurements of cluster shapes are based on detailed strong lensing analyses, from which we have derived shapes for 45 DM haloes in 39 galaxy clusters. Six of the 39 clusters have double peaks, for which we have measured shapes of individual DM peaks separately. In addition to DM haloes, we have also measured shapes of the BCG of each DM halo by diagonalizing the surface brightness tensor computed from F814W-band {\it HST} images. 

We have obtained the mean value of ellipticities of DM haloes, $\langle e_{\rm SL}\rangle = 0.482\pm0.028$, and those of BCGs, $\langle e_{\rm BCG}^{R_{ab}}\rangle=0.308\pm0.027$, $0.37\pm0.024$, and $0.421\pm0.026$ at $R_{ab}=10$, 20, and 30~pkpc, respectively. We have found that DM haloes are on average more elongated than BCGs with mean value of differences of their ellipticities of $\langle e_{\rm SL} - e_{\rm BCG}^{20}\rangle = 0.11\pm0.03$. The ellipticities of DM haloes and BCGs, and their differences do not strongly depend on the redshift. We have also found that orientations of DM haloes and BCGs are well aligned with each other and the degree of alignment is almost independent of the redshift. Mean values of the alignment angles are $\langle|\theta_{\rm SL}-\theta_{\rm BCG}^{R_{ab}}|\rangle=23.1\pm3.8$, $22.2\pm3.9$, and $23.3\pm3.3$~deg at $R_{ab}=10$, 20, and 30~pkpc, respectively. 

To interpret our observed results, we have computed projected shapes of DM haloes and CGs in the Horizon-AGN simulation. We have extracted 1265 DM haloes with FOF mass higher than $10^{12.5}M_{\odot}$ at $\langle z \rangle=0.39$ and created projected particle distributions. We regard three different projection directions as independent to obtain 3795 projected DM haloes in our analysis. Since in the Horizon-AGN simulation there is no halo whose mass scale is comparable to those of observed galaxy clusters, $M_{\rm vir}\sim10^{15}M_{\odot}$, we have focused on the mass dependence of shapes in the range of $10^{12.5}M_{\odot} < M_{\rm vir} < 10^{14.5}M_{\odot}$. We have computed ellipticities by a tensor method for DM haloes without substructure and for CGs with substructures for a fair comparison with observations. 

We have found that, for DM haloes, there is a clear trend that ellipticities become higher with increasing halo masses. Their mean values increase from $0.233$ at $M_{\rm vir} = 10^{12.6}M_{\odot}$ to $0.374$ at $10^{14.3}M_{\odot}$. Ellipticities of BCGs show the similar but weaker trend for the outer regions $R_{ab}=20$ and 30~pkpc, whereas ellipticities are almost constant against the host halo mass in the inner region $R_{ab}=10$~pkpc. Putting together, we have found that differences of ellipticities between DM haloes and CGs almost vanish on average. We have also found that the DM haloes and the CGs are well aligned with each other and the degree of the alignment exhibits the mass dependence such that the alignment becomes tighter with increasing halo masses. Mean values change from $\langle|\theta_{\rm DM}-\theta_{\rm CG}^{20}|\rangle = 30.6\pm0.6$~deg at $M_{\rm vir} = 10^{12.6}M_{\odot}$ to $18.3\pm5.5$~deg at $10^{14.3}M_{\odot}$. For all mass bins, the inner region of CGs shows tighter alignment than outer regions.

One of our main results is that observed values of the difference between ellipticities of DM haloes and BCGs, $\langle e_{\rm SL} - e_{\rm BCG}\rangle=0.11\pm0.03$, are on average larger than zero, which differs from the result of the Horizon-AGN simulation for which the average difference is consistent with zero. We note that our results appear to be consistent with \citet{2005ApJ...618..195G}, who find that position angles of intracluster light (ICL) distributions tend to be aligned well with those of BCGs and ICL distributions are more elongated than BCGs, if we assume that ICL distributions trace DM distributions as suggested by e.g., \citet{2019MNRAS.482.2838M}. We have discussed possible reasons for the difference between observations and the Horizon-AGN simulation in Section~\ref{sec:discussions_comparison_obs}. To discriminate different scenarios, however, future studies in both observations and simulations are needed. For the simulation side, larger box sizes are required so as to include higher mass haloes and also an exploration of baryon physics possibly to improve it. For the observational side, future large surveys such as the {\it Subaru} Hyper Suprime-Cam (HSC) \footnote[1]{https://hsc.mtk.nao.ac.jp/ssp/} \citep[e.g.,][]{2018PASJ...70S...1M, 2018PASJ...70S..27M, 2018PASJ...70S..20O, 2018PASJ...70S..25M} and the Large Synoptic Survey Telescope (LSST) \footnote[2]{https://www.lsst.org/} \citep[e.g.,][]{2009arXiv0912.0201L, 2019ApJ...873..111I}, as well as deep imaging by space telescopes such as the James Webb Space Telescope (JWST) \footnote[3]{https://www.jwst.nasa.gov/} \citep[e.g.,][]{2006SSRv..123..485G}, the Wide Field Infrared Survey Telescope (WFIRST) \footnote[4]{https://wfirst.gsfc.nasa.gov/}, the {\it Euclid} \footnote[5]{https://sci.esa.int/web/euclid/} would help to extend samples of strong lensing clusters and improve strong lensing constraints for individual clusters.

\section*{Acknowledgements}
We would like to thank Masahiro Takada and Massimo Meneghetti for useful discussions and comments. We also thank the anonymous referee for useful comments and suggestions.
T.O. is supported by Advanced Leading Graduate Course for Photon Science (ALPS) at the University of Tokyo. This work is supported partly by JSPS Core-to-Core Program ``International Network of Planetary Sciences''. This work is supported in part by Japan Society for the Promotion of Science (JSPS) KAKENHI Grant Number JP17J05056 (T.O.), JP18K03693 (M.O.), JP15H05892 (M.O.), JP17K14273 (T.N.), JP18K03704 (T.K.), and JP18H01247 (Y.S.). T.N. acknowledges Japan Science and Technology Agency (JST) CREST Grant Number JPMJCR1414.


\bibliographystyle{mnras}
\bibliography{reference} 

\appendix

\section{Strong lens mass models}
\label{sec:strong_lensing_method}

Strong lens mass models that are used in the analysis of this paper are summarized in Table~\ref{tab:slens}. All the mass models are constructed using the software {\sc glafic} \citep{2010PASJ...62.1017O}. The mass models of HFF clusters have already been presented in \citet{2016ApJ...819..114K} and \citet{2018ApJ...855....4K}, whereas those of CLASH and RELICS clusters have not been published elsewhere.

We follow \citet{2016ApJ...819..114K} for mass modeling procedure of CLASH and RELICS clusters. We assume simply parametrized mass models that consist of halo components modeled by an elliptical NFW profile and cluster galaxies modeled by an elliptical pseudo-Jaffe profile. To reduce the number of parameters, we assume scaling relations between galaxy luminosities and model parameters (velocity dispersions and truncation radii) of the pseudo-Jaffe profile. Ellipticities and position angles of cluster member galaxies are fixed to measured values of their light profiles, whereas ellipticities and position angles of halo components are treated as free parameters. We may also add external perturbations to the lens potential. We start with a simple mass model, and keep adding more halo components or external perturbations until we obtain reasonably good fit. Interested readers are referred to \citet{2016ApJ...819..114K} for more details.

We optimize model parameters so that the model can reproduce positions of multiple images. We rely on previous work as listed in Table~\ref{tab:slens} for identifications of multiple images and spectroscopic redshift information for some of them. Positional uncertainties of multiple images are set so as to achieve reasonably good fit i.e., reduced $\chi^2$ being of order one. $\chi^2$ is defined by differences of observed and model-predicted image positions evaluated in the source plane \citep[see Appendix~2 of][for more details]{2010PASJ...62.1017O}. The minimum $\chi^2$ for our best-fitting models are listed in Table~\ref{tab:slens}. Errors of model parameters are estimated using the Markov chain Monte Carlo method.

\begin{table*}
\centering
\caption{Summary of strong lens mass modeling using {\sc glafic} \citep{2010PASJ...62.1017O}. $N_{\rm sys}$ denotes the number of multiple image systems, $N_{\rm sys,spec}$ is the number of multiple image systems with spectroscopic redshifts, and $N_{\rm img}$ is the total number of multiple images used for mass modeling. The assumed positional error of multiple images in the image plane is shown by $\sigma_{\rm img}$. The minimum $\chi^2$ and degree of freedom are indicated by $\chi^2_{\rm min}$ and dof, respectively. }
\label{tab:slens}
\begin{tabular}{llcccccl} 
\hline
Survey & Cluster name & $N_{\rm sys}$ & $N_{\rm sys, spec}$ & $N_{\rm img}$ 
& $\sigma_{\rm img}$ [${}''$] & $\chi^2_{\rm min}$/dof & References \\
\hline
HFF    & Abell~2744           & 45 & 24 & 132 & 0.4 & 130.2/134 & 1 \\ 
HFF    & MACSJ0416.1$-$2403   & 75 & 34 & 202 & 0.4 & 240.0/196 & 1 \\ 
HFF    & MACSJ1149.5+2223     & 36 & 16 & 108 & 0.4 & 100.1/103 & 2 \\ 
HFF    & Abell~S1063          & 53 & 19 & 141 & 0.4 & 136.2/138 & 1 \\ 
CLASH  & Abell~209            &  3 &  0 &   7 & 0.8 & 2.8/1   & 3,4 \\ 
CLASH  & Abell~383            &  8 &  6 &  23 & 0.4 & 22.5/18  & 3,4 \\ 
CLASH  & MACSJ0329.7$-$0211   &  9 &  8 &  23 & 0.4 & 16.3/12  & 5,4 \\ 
CLASH  & MACSJ0429.6$-$0253   &  3 &  2 &  11 & 0.4 & 7.2/9   & 5,4 \\ 
CLASH  & MACSJ0744.9+3927     & 10 &  0 &  25 & 0.4 & 6.7/8   & 3,4 \\ 
CLASH  & Abell~611            &  3 &  2 &  14 & 0.4 & 11.6/12  & 3,4 \\ 
CLASH  & MACSJ1115.9+0129     &  3 &  1 &   9 & 0.6 & 4.9/3   & 5,4 \\ 
CLASH  & MACSJ1206.2-0847     & 27 & 27 &  82 & 0.4 & 79.9/83  & 6,4 \\ 
CLASH  & MACSJ1311.0$-$0310   &  3 &  1 &   8 & 0.6 & 7.2/4   & 5,4 \\ 
CLASH  & RXJ1347.5$-$1145     &  8 &  4 &  20 & 0.4 & 1.9/4   & 7,5,4 \\ 
CLASH  & MACSJ1423.8+2404     &  3 &  2 &  12 & 0.8 & 6.9/9   & 3,4 \\ 
CLASH  & MACSJ1720.3+3536     &  7 &  0 &  22 & 0.6 & 16.1/14  & 3,4 \\ 
CLASH  & Abell~2261           & 11 &  0 &  28 & 0.4 & 13.4/13  & 3,4 \\ 
CLASH  & MACSJ1931.8$-$2635   &  7 &  7 &  19 & 0.4 & 17.9/12  & 5,4 \\ 
CLASH  & RXJ2129.7+0005       &  7 &  7 &  22 & 0.4 & 17.1/21  & 5,4 \\ 
CLASH  & MS2137$-$2353        &  3 &  3 &  10 & 0.6 & 5.7/6   & 3,4 \\ 
CLASH  & MACSJ0647.8+7015     & 11 &  0 &  31 & 0.4 & 24.3/20  & 3,4 \\ 
CLASH  & MACSJ2129.4$-$0741   & 11 & 11 &  38 & 0.6 & 45.6/37  & 5,4 \\ 
RELICS & Abell~2163           &  4 &  0 &  15 & 0.4 & 6.6/12  & 8,4 \\ 
RELICS & Abell~2537           &  8 &  1 &  29 & 0.6 & 16.1/23  & 8,4 \\ 
RELICS & Abell~3192           &  5 &  2 &  16 & 0.8 & 7.4/6   & 9,4 \\ 
RELICS & Abell~697            &  3 &  0 &   9 & 0.4 & 6.7/6   & 10,4 \\ 
RELICS & Abell~S295           &  6 &  4 &  18 & 0.4 & 5.4/13  & 10,4 \\ 
RELICS & ACT-CL~J0102$-$49151 & 10 &  0 &  28 & 0.6 & 17.6/15  & 8,4 \\ 
RELICS & CL~J0152.7$-$1357    &  8 &  1 &  24 & 0.4 & 8.1/16  & 11,4 \\ 
RELICS & MACSJ0159.8$-$0849   &  4 &  0 &  10 & 0.6 & 5.6/4   & 10,4 \\ 
RELICS & MACSJ0257.1$-$2325   &  4 &  0 &  12 & 0.4 & 10.1/7   & 12,4 \\ 
RELICS & MACSJ0308.9+2645     &  3 &  0 &   7 & 0.4 & 0.7/1   & 13,4 \\ 
RELICS & MACSJ0417.5$-$1154   & 20 &  7 &  54 & 0.4 & 29.4/40  & 14,4 \\ 
RELICS & MACSJ0553.4$-$3342   & 10 &  2 &  30 & 0.8 & 29.9/25  & 15,4 \\ 
RELICS & PLCK~G171.9$-$40.7   &  5 &  0 &  16 & 0.4 & 11.7/7   & 13,4 \\ 
RELICS & PLCK~G308.3$-$20.2   & 11 &  0 &  31 & 0.6 & 17.8/18  & 16,4 \\ 
RELICS & RXC~J0142.9+4438     &  4 &  0 &  14 & 0.4 & 8.8/9   & 8,4 \\ 
RELICS & RXC~J2211.7$-$0350   &  3 &  1 &  11 & 0.4 & 2.7/3   & 8,4 \\ 
RELICS & SPT-CL~J0615$-$5746  &  6 &  5 &  22 & 0.4 & 5.2/17  & 17,4 \\ 
\hline
\end{tabular}
\flushleft{References -- 
(1) \citet{2018ApJ...855....4K}; 
(2) \citet{2016ApJ...819..114K};
(3) \citet{2015ApJ...801...44Z};
(4) this paper;
(5) \citet{2019arXiv190305103C};
(6) \citet{2017A&A...607A..93C};
(7) \citet{2018ApJ...866...48U};
(8) \citet{2018ApJ...859..159C};
(9) \citet{2013MNRAS.429..833H};
(10) \citet{2018ApJ...863..145C};
(11) \citet{2019ApJ...874..132A};
(12) \citet{2011MNRAS.410.1939Z};
(13) \citet{2018ApJ...858...42A};
(14) \citet{2019ApJ...873...96M};
(15) \citet{2017MNRAS.471.3305E};
(16) \citet{2017ApJ...839L..11Z};
(17) \citet{2018ApJ...863..154P}.
}
\end{table*}


\bsp	
\label{lastpage}
\end{document}